\documentclass[aps,pra,superscriptaddress,twocolumn]{revtex4-2}
\usepackage{bm}
\usepackage{amsmath}
\usepackage{amsthm}
\usepackage{amsfonts}
\usepackage{amssymb}
\usepackage{latexsym}
\usepackage[utf8]{inputenc}
\allowdisplaybreaks

\usepackage{slashed}
\usepackage{float}
\usepackage{subcaption}
\usepackage{dcolumn}
\usepackage{graphicx}

\newcommand{\half}{\textstyle{\frac12}}

\def\LZasquared{{\ln\left[\half(Z\alpha)^{-2}\right]}}
\newcommand{\LZasquaredmmu}{{\ln\left[\frac{m_1}{2\,\mu}(Z\alpha)^{-2}\right]}}
\newcommand{\LZasquaredpos}{{\ln\left[(Z\alpha)^{-2}\right]}}
\newcommand{\LZasquaredeta}{{\ln\left[\frac{1}{2\,\eta_1}(Z\alpha)^{-2}\right]}}

\begin{document}

\title{Radiative corrections of the order $\bm{\alpha\,(Z\,\alpha)^6}$ for rotational states of  two-body systems}

\author{Vojt\v{e}ch Patk\'o\v{s}}
\affiliation{Faculty of Mathematics and Physics, Charles University,  Ke Karlovu 3, 121 16 Prague
2, Czech Republic}

\author{Krzysztof Pachucki}
\affiliation{Faculty of Physics, University of Warsaw,
             Pasteura 5, 02-093 Warsaw, Poland}

\begin{abstract}
The analytical calculation of the complete $\alpha\,(Z\,\alpha)^6$ one-loop radiative correction to energies of two-body systems with
the angular momenta $l>0$, consisting of a pointlike particle and an extended-size nucleus
with arbitrary masses and spin 1/2, is presented.
The obtained results apply to a wide variety of two-body systems, such as hydrogen, muonium, positronium, and  antiprotonic atoms.
\end{abstract}
\maketitle

\section{Introduction}
Hadronic two-body systems, such as antiprotonic atoms in circular states $l\sim n$, give the possibility to probe the existence
of the long-range  interactions between hadrons, which is not possible by other means. 
The emission spectroscopy of light antiprotonic atoms is feasible at CERN \cite{nancypaul}, and from the theoretical side these atoms can be very accurately calculated.
In fact, in a highly excited circular state the effective coupling $Z\,\alpha/n$ is much smaller than one, and so the  nonrelativistic QED (NRQED) 
approach can be used to obtain the  energy levels even for high Z-nuclei.
Such calculations for an arbitrary mass ratio and arbitrary state  up to the order $(Z\,\alpha)^6$ were recently performed in Refs. \cite{zatorski:22, patkos:24:twobodyP},
and here we extend this result to the orders $\alpha\,(Z\,\alpha)^6$ and $Z^2\,\alpha\,(Z\,\alpha)^6$. 

Other two-body systems, such as hydrogen and hydrogen-like ions, serve for determination of the fundamental physical constants \cite{codata:22},
because they can be measured and calculated with high accuracy. Significant progress has been achieved in recent years by the inclusion
of the nuclear charge radii obtained from muonic hydrogen and other light muonic atoms \cite{muH1, muH2, muD, mualpha, muhelion, rmp:2024}. 
The current value of the Rydberg constant,
based mainly on the precisely measured $1S-2S$ transition in H \cite{1S2S} and $2S-2P$ in $\mu$H \cite{muH1,muH2},
has a relative accuracy of  $1.1\cdot10^{-12}$, limited by uncertainties in theoretical predictions for H and $\mu$H \cite{codata:22}. 
These uncertainties mainly  come from the two-loop electron self-energy, the radiative recoil, and nuclear polarizability in the case of muonic atoms. 
The radiative recoil correction is a topic of this work. 

In this paper,  employing units $\hbar = c = 1$ and $e^2 = 4\,\pi\,\alpha$, we perform a calculation at the $\alpha\,(Z\,\alpha)^6$ 
order for two-body systems with arbitrary masses, including self-energy of an orbiting particle and with an arbitrary nucleus.
In the first step, we consider the states with $l>0$.
The lower-order terms were recently obtained for $l=0$ states in Ref.~\cite{adkins:23}, and for 
$l>0$ in Refs.~\cite{zatorski:22,patkos:24:twobodyP}. The $\alpha^7$ corrections are currently known only in the nonrecoil limit \cite{jentschura:05}, 
and here we derive them for an arbitrary mass ratio. The results obtained may also find  applications  in more complicated few-electron systems such as the helium atom, 
where discrepancies between theoretical predictions and experimental values for the ionization energies have been observed  \cite{He1, He2,He3},
and they might come from a similar calculation of radiative $\alpha^7\,m$ correction for triplet states of the He atom \cite{patkos:21:rad}.

\section{Radiative $\alpha\,(Z\,\alpha)^6$ correction}
The radiative (electron self-energy) $\alpha\,(Z\,\alpha)^6$ correction to energy $E^{(7)}_\mathrm{rad}$ of a two-body system 
can be expressed as a combination of terms with all possible spin couplings \cite{zatorski:22},
\begin{align}
E^{(7)}_\mathrm{rad} = &\ \frac{\mu\,\alpha(Z\alpha)^6}{\pi}\big\langle \mathcal{E}_\mathrm{NS} + \vec L\cdot\vec s_1\,\mathcal{E}_\mathrm{S1}
+ \vec L\cdot\vec s_2\,\mathcal{E}_\mathrm{S2} \nonumber \\
&\ + \vec s_1\cdot\vec s_2\,\mathcal{E}_\mathrm{SS}
+ (L^i L^j)^{(2)}\,s_1^i s_2^j\,\mathcal{E}_\mathrm{LL}
\big\rangle  \,, \label{01}
\end{align}
where $\mu$ is the reduced mass, $Z\,\alpha = -e_1\,e_2/(4\,\pi)$, $Z$ is the charge number of the nucleus which has a particle number $2$,
$\vec s_i$ is the spin of the $i$-th particle, and
\begin{equation}
(L^i L^j)^{(2)} = \frac12\,(L^i L^j + L^j L^i) - \frac{\delta^{ij}}{3}\,\vec L^2\,,
\end{equation}
which is a symmetric traceless tensor. The coefficients ${\cal E}_X$ in Eq. (\ref{01}) with different X are functions of 
the principal quantum number $n$ and the angular momentum $l$,  and their calculation is the subject of this work.
 
At first $E^{(7)}_\mathrm{rad} $  is divided into three parts,
\begin{align}
E^{(7)}_\mathrm{rad} = E_L + E_M + E_H\,, \label{03}
\end{align}
where the low-energy part $E_L$ corresponds to the frequency of the radiative photon $\omega \sim m_1\,\alpha^2$,
the middle-energy part $E_M$ comes from the region of $\omega \sim m_1\,\alpha$,
and the high-energy part $E_H$ corresponds to $\omega \sim m_1$.
We will use dimensional regularization with $d=3-2\,\varepsilon$ to avoid divergences
and the $1/\varepsilon$ singularity will cancel out in the  sum in Eq. (\ref{03}).

\section{Low-energy part $E_L$}

The low-energy contribution of the order $\alpha\,(Z\alpha)^6$ is further divided into three parts,
\begin{align}
E_L = E_{L1} + E_{L2} + E_{L3} \label{02}\,.
\end{align}
These parts will be evaluated in the subsequent sections as corrections to the
leading low-energy contribution $E_{L0}$ of the order $\alpha\,(Z\alpha)^4$,
namely to the Bethe logarithm.

\subsection{$E_{L0}$}
Let us consider  the nonrelativistic Hamiltonian for a two-body system in $d$ dimensions,
\begin{align}\label{04}
H = &\ \frac{p_1^2}{2\,m_1} + \frac{p_2^2}{2\,m_2} + V(r)\,,\\
V(r) =&\ \frac{e_1\,e_2}{4\,\pi}\,\bigg[\frac{1}{r}\bigg]_\epsilon\,, \\
\bigg[\frac{1}{r}\bigg]_\epsilon =&\ m_1^{2\,\varepsilon} \int\frac{d^dk}{(2\,\pi)^d}\,\frac{4\,\pi}{k^2}\,,
\end{align}
where $\vec r = \vec r_1-\vec r_2$.
The leading nonrelativistic (dipole) low-energy contribution is
\begin{align}
E_{L0} =&\ \frac{e_1^2}{m_1^{2-2\,\varepsilon}}\int \frac{d^d k}{(2\,\pi)^d\,2\,k}\,
\left(\delta^{ij}-\frac{k^i\,k^j}{k^2}\right)\,
\nonumber \\ &\times
\left< \phi \left|p^i_1\,\frac{1}{E-H-k}\,p^j_1 \right|\phi\right> \,,
\label{05}
\end{align}
where $H$ is the nonrelativistic Hamiltonian in $d$ dimensions from Eq.~(\ref{04}).
The wave function $\phi$ denotes
the nonrelativistic Schr\"{o}dinger--Pauli wave function  in the center of mass frame ($\vec p_1 = - \vec p_2 = \vec p$\,).
In the following, we will denote the expectation value
of an arbitrary operator $Q$,
evaluated with the nonrelativistic Schr\"{o}dinger--Pauli
wave function, by the shorthand notation $\langle Q \rangle$.

After the $d$-dimensional
integration with respect to $k$, and the expansion in
$\varepsilon$, $E_{L0}$ becomes
\begin{align}
\label{EL0intermediate}
E_{L0} =&\ (4\,\pi)^\varepsilon\,\Gamma(1+\varepsilon)\,
\frac{2\,\alpha}{3\,\pi\,m_1^2}\,
\bigg< \vec p_1\,(H-E)\bigg\{\frac{1}{2\,\varepsilon}+\frac{5}{6} \nonumber \\
&\ -
\ln\left[\frac{2(H-E)}{m_1}\right]
\bigg\}\vec p_1\,\bigg> \,,
\end{align}
where we ignore terms of order $\varepsilon$ and higher.
The factor $(4\,\pi)^\varepsilon\,\Gamma(1+\varepsilon)$
appears in all the terms, and thus we will omit consistently in all matrix elements.
The contribution $E_{L0}$ can thus be rewritten as
\begin{eqnarray}
\label{EL0}
E_{L0} &=&
\frac{4\,\alpha}{3\,m_1^2}\, Z\,\alpha\,
\left\{ \frac{1}{2\,\varepsilon} + \frac{5}{6} + \ln\left[\frac{m_1}{\mu\,(Z\,\alpha)^{2}}\right]
\right\} \, \langle \delta^d(r) \rangle
\nonumber \\ &&
- \frac{2\,\alpha}{3\,\pi\,m_1^2} \left< \vec p\,(H-E) \,
\ln\left[\frac{2(H-E)}{\mu\,(Z\,\alpha)^2}\right] \,
\vec p\,\right>\,,
\end{eqnarray}
where the last term is the so-called Bethe logarithm \cite{bethe}.
\subsection{$E_{L1}$}
We consider now all possible relativistic corrections to Eq.~(\ref{EL0})
and introduce the notation
\begin{align}
\label{deltaQ}
&\delta_Q\,\left< p^i\,\frac{1}{E-H-k}\,p^j \right> \equiv
\biggl< p^i\,\frac{1}{E-H-k}\,(Q-\langle Q\rangle)\,
\nonumber \\ &
\frac{1}{E-H-k}\,p^j \biggr> +
2\, \left< Q\,\frac{1}{(E-H)'}\,p^i\,\frac{1}{E-H-k}\,p^j \right>\,,
\end{align}
where $Q$ is an arbitrary operator. $\delta_Q$ involves the
first-order perturbations
to the Hamiltonian, to the energy, and to the wave function.
The correction $E_{L1}$ is the perturbation of $E_{L0}$
by the relativistic Breit Hamiltonian $H^{(4)}$,
which in $d$ dimensions is (setting $e_1=-e$, $e_2 = Z e$)
\begin{align}\label{eq:hbreit}
 H^{(4)}  = &\ H'^{(4)} + H''^{(4)}\,, \\
\label{eq:hbreit1}
 H'^{(4)} = &\ 
 \sum_{a=1,2}\biggl\{ -\frac{p_a^4}{8\,m_a^3} +\frac{\pi\,Z\alpha}{2}\bigg[\frac{1}{m_a^2} + \frac43\,r_{Ea}^2\bigg]\,\delta^d(r)\biggr\}
\nonumber \\ &\
+\frac{Z\alpha}{2\,m_1\,m_2}\,p_1^i\,
\biggl[\frac{\delta^{ij}}{r}+\frac{r^i\,r^j}{r^3}
\biggr]_\epsilon\,p_2^j \nonumber \\
&\ + \frac{g_1\,g_2\,\pi\,Z\alpha}{4\,d\,m_1\,m_2}\,\sigma_1^{ij}\,\sigma_2^{ij}\,\delta^d(r)\,,
  \\
 H''^{(4)}  = &\
 \sum_{a=1,2} \frac{g_a-1}{4\,m_a^2}\,\sigma_a^{ij}\,\bigl(\nabla^i_a V\bigr) p_a^j
 \nonumber \\ &\
+\frac{g_1\,g_2}{16\,m_1\,m_2}\,\sigma_1^{ik}\,\sigma_2^{jk}\,
\left(\nabla^i\,\nabla^j-\frac{\delta^{ij}}{d}\,\nabla^2\right)
V \nonumber \\ &\
-\frac{1}{4\,m_1\,m_2}\,\bigl(\nabla^i V\bigr) \biggl(g_1\,\sigma_1^{ij}\,p^j_2
-g_2\,\sigma_2^{ij}\,p^j_1 \biggr)\,,
\label{eq:hbreit2}
\end{align}
where  $r_{E}^2$ is the mean square charge radius,
$\delta^{d}(r)$ is the Dirac $\delta$-function in $d$ dimensions,
and $\sigma^{ij} = [\sigma^i\,,\,\sigma^j]/(2\,i)$.
In $d=3$ spatial dimensions, the
matrices $\sigma^{ij}$ reduce to $\sigma^{ij}=\epsilon^{ijk} \, \sigma^k$,
and the Breit Hamiltonian in the center of mass frame becomes
\begin{widetext}
\begin{align}
H^{(4)} =&\ -\frac{\vec p^{\,4}}{8\,m_1^3}\  -\frac{\vec p^{\,4}}{8\,m_2^3}
-Z\alpha \,\biggl\{
\frac{1}{2\,m_1\,m_2}\,p^i\,
\biggl(\frac{\delta^{ij}}{r}+\frac{r^i\,r^j}{r^3}\biggr)\,p^j
+\frac{g_1\,g_2}{4\,m_1\,m_2}\,\biggl[\frac{s_1^i\,s_2^j}
{r^3}\,\biggl(\delta^{ij}-3\,\frac{r^i\,r^j}{r^2}\biggr)
-\frac{8\,\pi}{3}\,\vec s_1\cdot\vec s_2\,\delta^{3}(\vec r)\biggr]
\nonumber \\ &\
-\frac{\vec r\times\vec p}{2\,r^3} \cdot\biggl[
\frac{g_1}{m_1\,m_2}\,\vec s_1
+\frac{g_2}{m_1\,m_2}\,\vec s_2
+\frac{(g_2-1)}{m_2^2}\,\vec s_2
+\frac{(g_1-1)}{m_1^2}\,\vec s_1\biggr]\biggr\}
 + \frac{2\,Z\alpha}{3} \biggl( \frac{3}{4\,m_1^2}+ \frac{3}{4\,m_2^2} +r_{E1}^2+  r_{E2}^2 \biggr) \, \pi\,\delta^{3}(\vec{r})\,. \label{Breitgeneral}
\end{align}
\end{widetext}
We will use this $d=3$ form of $H^{(4)}$ also later in the calculation of the second-order correction.
Additionaly, we note that the first particle is point-like, so $r_{E1}^2=0$ and $g_1=2$.
The second particle will be considered with finite nuclear size, and we will calculate the radiative corrections
only for the first particle. However, in the case of antiprotonic atoms  we will drop these assumptions for the first particle 
and include radiative corrections for the second particle in Sec. VIII.

We now split $E_{L1}$ by introducing an intermediate cutoff $\Lambda$
\begin{eqnarray}
E_{L1} &=& \frac{e^2}{m_1^{2-2\,\varepsilon}}\left(\int_0^\Lambda + \int_\Lambda^\infty\right)
 \frac{d^d k}{(2\,\pi)^d\,2\,k}\,
\left(\delta^{ij}-\frac{k^i\,k^j}{k^2}\right)\,
\nonumber \\ && \times
\delta _{H^{(4)}}\left< p_1^i\,\frac{1}{E-H-k}\,p_1^j \right>\,.
\end{eqnarray}

After the $Z\,\alpha$ expansion with $\Lambda = \lambda\,(Z\,\alpha)^2$,
one goes subsequently to the limits  $\varepsilon\rightarrow 0$
and $\lambda\rightarrow \infty$.
Under the assumption that $l\neq0$,
we may perform an expansion in $1/k$ in the second part and obtain
\begin{align}
E_{L1} = & \
\frac{2\,\alpha}{3\,\pi\,m_1^2}\,\int_0^\Lambda dk\,k\,
\delta_{H^{(4)}}\left< \,\vec p_1\,\frac{1}{E-H-k}\,\vec p_1\,\right>
\nonumber \\ &
+ \frac{\alpha}{3\,\pi\,m_1^{2-2\,\varepsilon}}\,
\left[1+\varepsilon\,\left(\frac{5}{3}-2\,\ln2 \right)\right]
\int_\Lambda^\infty
dk\,\frac{1}{k^{1+2\,\varepsilon}}\,
\nonumber \\
& \times\bigg\{2\, \left< 
H^{(4)}\,\frac{1}{(E-H)'}\,
[\,\vec p_1,[\,H,\vec p_1\,]]\right> \nonumber \\
&\ + \bigg<[\,\vec p_1,[\,H^{(4)},\vec p_1\,]]\bigg> \bigg\} \,.
\end{align}
The second-order contribution in braces will vanish for states with $l\neq 0$.
In the calculations, we keep $g_2$ and $r_{E2}^2$ arbitrary.
After performing the $k$-integration and with the help
of commutator relations, it reads
\begin{widetext}
\begin{align}
\label{EL1intermediate}
E_{L1} =&\
\frac{\alpha}{\pi}\,\frac{(Z\,\alpha)^6}{n^3}\,\mu\,\beta_1 +
\frac{\alpha}{3\,\pi}\,
\left\{\frac{1}{2\,\varepsilon}+\frac{5}{6} +
\LZasquaredmmu\right\}\,
\frac{1}{m_1^2} \bigg<\vec p\,4\pi\,\delta^d(r)\,\vec p\, 
\nonumber \\ &\ 
\times\bigg[\frac{Z\alpha}{4}\bigg(\frac{1}{m_1^2}+\frac{1}{m_2^2} + \frac43\,r_{E2}^2\bigg) 
+\frac{Z\alpha}{m_1\,m_2}\bigg(\frac23 - \frac29\varepsilon\bigg)
+\frac{g_2\,Z\alpha}{4\,m_1\,m_2}\bigg(\frac13 + \frac29\varepsilon\bigg)\,\sigma_1^{ij}\,\sigma_2^{ij}\bigg]
\nonumber \\ &\ 
-\frac{Z\alpha}{m_1\,m_2}\,p^i\,\bigg(\frac{\delta^{ij}}{r^3} - 3\frac{r^i \,r^j}{r^5}\bigg)_\epsilon\,p^j
+\frac{g_2\,Z\alpha}{4\,m_1\,m_2}\,\sigma_1^{ik}\,\sigma_2^{jk}\,\big[p^i\,4\pi\,\delta^d(r)\,p^j\big]^{(2)}
\nonumber \\ &\
+{\rm i}\,\frac{Z\alpha}{2}\bigg(\frac{\sigma_1^{ij}}{2\,m_1^2} + \frac{(g_2-1)\,\sigma_2^{ij}}{2\,m_2^2}
 + \frac{2\,\sigma_1^{ij}+g_2\,\sigma_2^{ij}}{2\,m_1\,m_2}\bigg)\,p^i\,4\pi\,\delta^d(r)\,p^j\bigg>\,,
\end{align}
where the expectation value is expressed in the center of mass system.
Here, $\beta_1$ is
a dimensionless quantity, defined as a finite part of the $k$-integral
with divergent terms proportional to $\lambda^n$
($n = 1, 2, \dots$) and  $\ln(\lambda/\mu)$ in the
limit of large $\lambda$ omitted,
\begin{align}
\label{defbeta1}
\frac{\alpha}{\pi}\,\frac{(Z\,\alpha)^6}{n^3}\,\beta_1 = &\
\lim_{\lambda\rightarrow\infty}
\frac{2\,\alpha}{3\,\pi\,m_1^2\,\mu}\,
\int_0^\Lambda dk\,k \,
\delta_{H^{(4)}}\,\left< p_1^i\,\frac{1}{E-H-k}\,p_1^i \right> \nonumber \\
 = &\ \frac{\alpha}{\pi}\,\frac{(Z\,\alpha)^6}{n^3}\,\bigg[\beta^\mathrm{NS}_1 + \vec L\cdot\vec s_1\,\beta_1^\mathrm{S1}
+\vec L\cdot\vec s_2\,\beta_1^\mathrm{S2} 
+ \vec s_1\cdot\vec s_2\,\beta_1^\mathrm{SS} + (L^i L^j)^{(2)} s_1^i s_2^j\,\beta_1^\mathrm{LL}\bigg]\,.
\end{align}
\end{widetext}
In all integrals with an upper limit $\Lambda$, to be discussed in the following,
the divergent terms in $\lambda$ will be subtracted.
In particular, the terms proportional
to $\ln(\lambda/\mu)$ but not $\ln(2\,\lambda/\mu)$ are subtracted, which leads to the presence of factor
$\half$ under the logarithm in Eq.~(\ref{EL1intermediate}).

\subsection{$E_{L2}$}
The second relativistic correction, $E_{L2}$, is the
nonrelativistic quadrupole contribution.
Specifically, it comes from  the quadratic in $k$ term
from the expansion of $\exp({\rm i}\,\vec k\cdot\vec r)$,
\begin{eqnarray}
E_{L2} &=& \frac{e^2}{m_1^{2-2\,\varepsilon}}\int \frac{d^d k}{(2\,\pi)^d\,2\,k}\,
\left(\delta^{ij}-\frac{k^i\,k^j}{k^2}\right)\,
\nonumber \\ && \times
\biggl[ \left< p_1^i\,({\rm i}\,\vec k\cdot\vec r_1)\,
\frac{1}{E-H-k}\,p_1^j
\,(-{\rm i}\,\vec k\cdot\vec r_1)\right>
\nonumber \\ &&
+\left< p_1^i\,({\rm i}\,\vec k\cdot\vec r_1)^2\,
\frac{1}{E-H-k}\,p_1^j\right> \biggr]
\,.
\end{eqnarray}
In a similar way as for $E_{L1}$, we split the integration into two parts,
by introducing a cutoff $\Lambda$. In the first part, with the
$k$-integral from $0$ to $\Lambda$, one can set $d=3$ and extract the
logarithmic divergence.
In the second part, with the $k$-integral from $\Lambda$ to $\infty$,
we perform a $1/k$ expansion and employ commutator relations,
with the intent of moving the operator $H-E$ to the far left
or right where it vanishes when acting on the Schr\"{o}dinger--Pauli wave function.
In this second part it is advantageous, instead of directly expanding the exponentials, at first to use the identity
\begin{equation}
e^{i\vec k\cdot\vec r}\,f(\vec p\,)\,e^{-i\vec k\cdot\vec r} = f(\vec p - \vec k)\,.
\end{equation}
Thus, after expanding the resolvent in $1/k$, we get for the expression in the expectation value
\begin{align}
&\ \left< p_1^i\,e^{i\vec k\cdot\vec r_1}\,\frac{(H-E)^3}{k^4}\,p_1^i\,e^{-i\vec k\cdot\vec r_1} \right>
   \nonumber \\
= &\ \frac{1}{k^4} \bigg< p_1^i\,\bigg(H-E - \frac{\vec p_1\cdot\vec k}{m_1} + \frac{k^2}{2\,m_1}\bigg)^3\,p_1^i \bigg>\,.
\end{align}
We expand the bracket and take into account only terms quadratic in $k$, contributing at the order $\alpha^7$. This leads to
\begin{align}
E^\Lambda_{L2} = &\ \frac{e^2}{m_1^{2-2\,\varepsilon}}\int_\Lambda^\infty \frac{d^d k}{(2\,\pi)^d\,2\,k^5}\,
\left(\delta^{ij}-\frac{k^i\,k^j}{k^2}\right)\,
\nonumber \\ &\ \times
\bigg< p_1^i\,
\bigg[\frac{3}{2\,m_1}(H-E)^2\,k^2 + \frac{2}{m_1^2} (\vec p_1\cdot\vec k)^2\,(H-E)
\nonumber\\ &\
+ \frac{1}{m_1^2} (\vec p_1\cdot\vec k) (H-E) (\vec p_1\cdot\vec k) \bigg]\,p_1^j
\bigg>
\,.
\end{align}
We now pass to the center of mass system, and the resulting expression, after performing $k$ integration and expansion for small $\varepsilon$, is
\begin{align}
\label{El2intermediate}
E_{L2}  = &\
\frac{\alpha}{\pi}\,\frac{(Z\,\alpha)^6}{n^3}\,\mu\,\beta_2 +
\frac{\alpha}{\pi}\,\bigg<
(\vec{\nabla} V)^2\,
\bigg[\frac{1}{m_1^3}\bigg(\frac{1}{2\varepsilon}\nonumber \\
&\ +\frac{5}{6}  +\LZasquaredmmu\bigg) + \frac{\mu}{m_1^4}
\bigg(\frac{1}{6\varepsilon}+\frac{14}{45}
\nonumber \\ &\
+\frac13\,\LZasquaredmmu\bigg)\bigg]
+\frac{Z\alpha}{m_1^4} \,\vec p\,4\pi\,\delta^d(r)\,\vec p\,\nonumber \\
&\ \times\left[\frac{1}{20\,\varepsilon}+\frac{3}{25}
+\frac{1}{10}\, \LZasquaredmmu\right]
\bigg> \,.
\end{align}
Here, $\beta_2 = \beta_2^\mathrm{NS}$ is defined as the finite part of the integral
[see the discussion following Eq.~(\ref{defbeta1})]
\begin{align}
\label{defbeta2}
 \frac{\alpha}{\pi}\,\frac{(Z\,\alpha)^6}{n^3}&\ \beta_2  = 
\frac{4\,\pi\,\alpha}{m_1^2\,\mu}\,\,\lim_{\lambda\rightarrow\infty} \int_0^\Lambda\frac{d^3k}{(2\,\pi)^3\,2\,k}\,
\left(\delta^{ij}-\frac{k^i\,k^j}{k^2}\right)\,\nonumber \\
&\times \left\{\left< \,p_1^i\,({\rm i}\,\vec k\cdot\vec r_1)^2\,
\frac{1}{E-H-k}\,p_1^j\,\right> \right.
\nonumber \\ &\
\left. +\left< \,p_1^i\,({\rm i}\,\vec k\cdot\vec r_1)\,
\frac{1}{E-H-k}\,p_1^j\,(-{\rm i}\,\vec k\cdot\vec r_1)\,\right>
\right\} \,.
\end{align}

\subsection{$E_{L3}$}
The third contribution, $E_{L3}$, originates from the relativistic
corrections to the coupling of the electron to the electromagnetic field.
These corrections can be obtained  from the Hamiltonian in Eq.~(\ref{04}),
and they have the form of a correction to the current
\begin{align}
\delta j_1^i =  &\ {\rm i}\,[H^{(4)},r_1^i ]  \nonumber \\
= &\ -\frac{1}{2\,m_1^3}\,p_1^i\,\vec p_1^{\,2}
+\frac{Z\alpha}{2\,m_1\,m_2}\bigg[\frac{\delta^{ij}}{r} + \frac{r^i\,r^j}{r^3}\bigg]_\varepsilon p_2^j \nonumber \\
&\ +\frac{g_1-1}{4\,m_1^2}\,\sigma_1^{ji}\,\nabla^j V
+\frac{g_2}{4\,m_1\,m_2}\,\sigma_2^{ji}\,\nabla^j V\,,
\end{align}
with $H^{(4)}$ given in Eq. (\ref{eq:hbreit}), and we keep $g_1$ arbitrary for now.
The corresponding correction $E_{L3}$ is
\begin{align}
E_{L3} = &\ 2\,\frac{e^2}{m_1^{1-2\,\varepsilon}}\int \frac{d^d k}{(2\,\pi)^d\,2\,k}\,
\left(\delta^{ij}-\frac{k^i\,k^j}{k^2}\right)\,\nonumber \\
&\ \times \left< \delta j_1^i\frac{1}{E-H-k}\,p_1^j \right> \,.
\end{align}
We now perform an angular averaging of the matrix element to bring the correction $E_{L3}$ into the form
\begin{align}
E_{L3} =2\,\frac{e^2}{m_1^{1-2\,\varepsilon}}\frac{d-1}{d}\int \frac{d^d k}{(2\,\pi)^d\,2\,k}\,
\left< \delta j_1^i
\frac{1}{E-H-k}\,p_1^i \right> \,.
\end{align}
We again split this integral into two parts.  In the first part, where
$k<\Lambda$, one can approach the limit $d=3$.
In the second part, with $k>\Lambda$, one performs a $1/k$-expansion
and obtains
\begin{align}
E_{L3}^\Lambda = &\ \frac{\alpha}{m_1\,\pi}\bigg(\frac59 + \frac{1}{3\varepsilon} 
- \frac23\ln\bigg[\frac{2\Lambda}{m_1}\bigg] \bigg) \left\langle\left[\delta j_1^i,\left[V,p_1^i\right]\right]\right\rangle
\,.
\end{align}
The expectation value for states with angular momentum $l>0$ can be written as
\begin{align}
\left[\delta j_1^i,\left[V,p_1^i\right]\right] = &\ - \frac{1}{2\,m_1^3}\,\left[p_1^ip_1^2,\left[V,p_1^i\right]\right] \nonumber \\
&\ +\frac{Z\alpha}{2\,m_1\,m_2}\,\bigg[\frac{\delta^{ij}}{r} + \frac{r^i\,r^j}{r^3}\bigg]_\varepsilon\left(-\nabla^i\nabla^j V\right) \nonumber \\
= &\ \bigg( -\frac{\mu}{m_1^3}  + \frac{1 - \varepsilon}{m_1\,m_2}\,\bigg)(\vec\nabla V)^2\,,
\end{align}
where  we used the identity
\begin{align}
\bigg[\frac{Z\alpha}{2}\bigg(\frac{\delta^{ij}}{r} + \frac{r^i\,r^j}{r^3}\bigg)\bigg]_\varepsilon\left(-\nabla^i\nabla^j V\right)
=\big(1 - \varepsilon\big)(\vec\nabla V)^2\,,
\end{align}
which follows from evaluation of this expression in momentum representation in $d$ dimensions, namely
\begin{widetext}
\begin{align}
(\vec\nabla V(r))^2 &\ =  (4\,\pi\,Z\,\alpha)^2\,m_1^{4\,\varepsilon}\int \frac{d^d q}{(2\,\pi)^d}\,e^{i\,\vec q\cdot \vec r} 
 \int \frac{d^d k}{(2\,\pi)^d} \frac{\vec k\cdot(\vec k-\vec q)}{k^2\,(\vec k-\vec q)^2}
\nonumber \\ 
&\ = 
-(\pi\,Z\,\alpha)^2\,m_1^{4\,\varepsilon}
\int \frac{d^d q}{(2\,\pi)^d}\,e^{i\,\vec q\cdot \vec r} \, 4^{\varepsilon}\,q^{1 - 2\,\varepsilon}
\frac{\tan(\varepsilon\,\pi)}{\varepsilon\,\pi}\,,
\end{align}
and
\begin{align}
\bigg[\frac{Z\alpha}{2}\bigg(\frac{\delta^{ij}}{r} + \frac{r^i\,r^j}{r^3}\bigg)\bigg]_\varepsilon\left(-\nabla^i\nabla^j V\right)  =&\  
-(4\,\pi\,Z\,\alpha)^2\,m_1^{4\,\varepsilon}\int \frac{d^d q}{(2\,\pi)^d}\,e^{i\,\vec q\cdot \vec r} \int \frac{d^d k}{(2\,\pi)^d}\,
\biggl(\delta^{ij}-\frac{k^i\,k^j}{k^2}\biggr)\,\frac{1}{k^2}\, \frac{(q-k)^i\,(q-k)^j}{(q-k)^2}
\nonumber \\ =&\
-(\pi\,Z\,\alpha)^2\,m_1^{4\,\varepsilon}
\int \frac{d^d q}{(2\,\pi)^d}\,e^{i\,\vec q\cdot \vec r} \, 4^{\varepsilon}\,q^{1 - 2\,\varepsilon}\,
\frac{\tan(\varepsilon\,\pi)}{\varepsilon\,\pi}\,(1-\varepsilon)\,.
\end{align}
For $E_{L3}$ we finally obtain
\begin{align}
\label{EL3intermediate}
E_{L3} = &\
\frac{\alpha}{\pi}\,\frac{(Z\,\alpha)^6}{n^3}\,\mu\,\beta_3
-\frac{\alpha}{\pi}
\,\left[\frac{1}{3\,\varepsilon}+
\frac{5}{9}+ \frac23
\LZasquaredmmu\right]\,\left[\frac{\mu}{m_1^4}  - \frac{(1-\varepsilon)}{m_1^2\,m_2}\right]
\left<\bigl(\vec{\nabla} V\bigr)^2\right>\,,
\end{align}
where $\beta_3$ is the finite part of the integral (in the center of mass system)
\begin{align}
\label{defbeta3}
\frac{\alpha}{\pi}\,\frac{(Z\,\alpha)^6}{n^3}\,\beta_3 =&\ 
-\frac{4\,\alpha}{3\,\pi\,m_1\,\mu}\,\lim_{\lambda\rightarrow\infty}
\int_0^\Lambda dk\,k\,\bigg< \bigg[\frac{1}{2\,m_1^3}\,p^i\,p^2
 + \frac{Z\alpha}{m_1\,m_2}\bigg(\frac{\delta^{ij}}{r} + \frac{r^i\,r^j}{r^3}\bigg)\,p^j \nonumber\\ &\
+\epsilon^{ijk}\bigg(\frac{g_1-1}{4\,m_1^2}\,\sigma_1^{k} + \frac{g_2}{4\,m_1\,m_2}\,\sigma_2^{k}\bigg)\,\nabla^jV\bigg]
\,\frac{1}{E-H-k}\,p^i\,\bigg> \nonumber \\
= &\ \frac{\alpha}{\pi}\,\frac{(Z\,\alpha)^6}{n^3}\,\bigg[\beta_3^\mathrm{NS} + \vec L\cdot\vec s_1\,\beta_3^\mathrm{S1} + \vec L\cdot\vec s_2\,\beta_3^\mathrm{S2}\bigg] \,.
\end{align}
\end{widetext}
Now we make the transition $g_1\rightarrow2$, but in the case of antiprotonic atoms, discussed in Sec. VIII  we would keep $g_1$ arbitrary.
This completes the treatment of the low-energy part in Eq. (\ref{02}), and the complete Bethe-log-like contributions are
\begin{align}
\beta^\mathrm{NS} =&\ \beta_1^\mathrm{NS} + \beta_2^\mathrm{NS} + \beta_3^\mathrm{NS}, \\
\beta^\mathrm{S1} =&\ \beta_1^\mathrm{S1} +  \beta_3^\mathrm{S1}, \\
\beta^\mathrm{S2} =&\ \beta_1^\mathrm{S2} +  \beta_3^\mathrm{S2}, \\
\beta^\mathrm{SS} =&\ \beta_1^\mathrm{SS},  \\
\beta^\mathrm{LL} =&\ \beta_1^\mathrm{LL}. 
\end{align}

\section{Middle-energy part}
\begin{figure}[H]
\centering
\begin{subfigure}{0.5\textwidth}
\includegraphics[width=\linewidth]{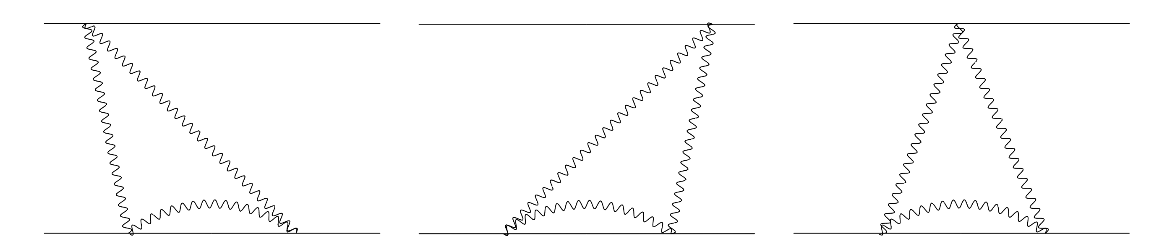}
\end{subfigure}
\caption{
 Time-ordered diagrams contributing to the middle-energy contribution $E_{M1}$.
\label{fig:Ei}}
\end{figure}

\begin{figure}[H]
\centering
\begin{subfigure}{0.5\textwidth}
\includegraphics[width=\linewidth]{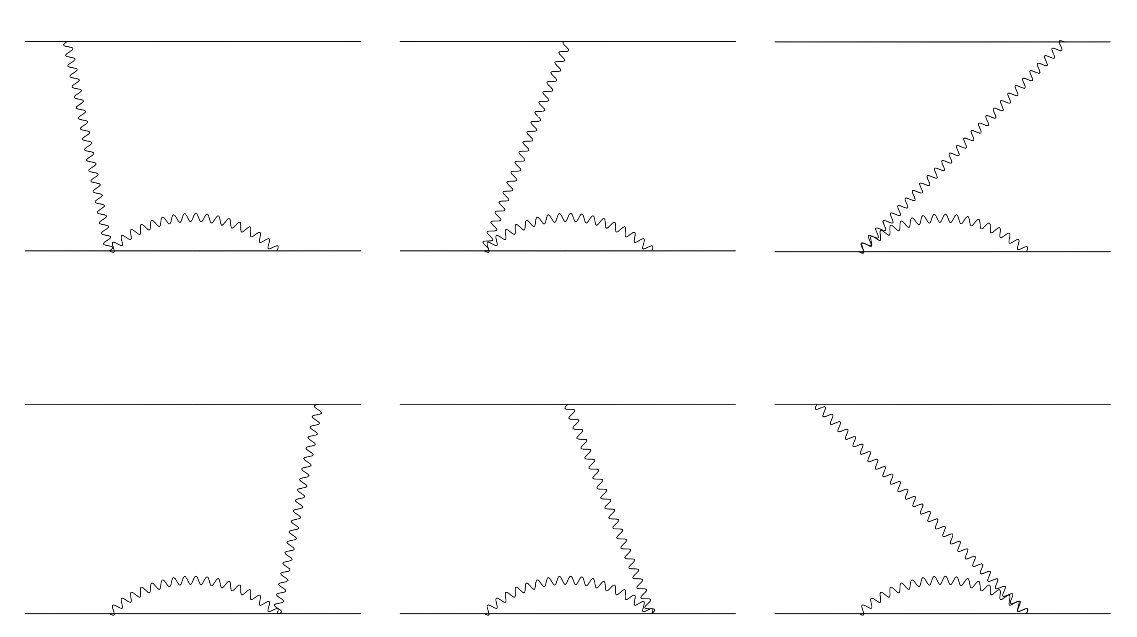}
\end{subfigure}
\caption{
Time-ordered diagrams contributing to the middle-energy contribution $E_{M2}$.
\label{fig:Ei2}}
\end{figure}

In the middle-energy part, the momenta of both the radiative and the exchanged photon are of the order $m_1\alpha$.
This part consists of two diagrams: the triple seagull contribution and a single seagull with retardation; see Figs.~\ref{fig:Ei} and~\ref{fig:Ei2}.
We follow the approach used in \cite{patkos:21:rad} for the case of two electrons and extend it to two particles with arbitrary masses.

\subsection{Triple seagull contribution}

The first middle-energy contribution is the triple seagull diagram given by Fig.~\ref{fig:Ei}, which is expressed 
(with $k_3$ being the radiative photon) as
\begin{widetext}
\begin{eqnarray}
E_{M1} &=& \frac{e^6\,Z^2}{m_1^{2-6\,\varepsilon}\,m_2}\int\frac{d^d k_1}{(2\pi)^d\,2k_1}
\int\frac{d^d k_2}{(2\pi)^d\,2k_2}\int\frac{d^d k_3}{(2\pi)^d\,2k_3}
\,\delta_\perp^{ik}(k_1)\delta_\perp^{jk}(k_2)\delta_\perp^{ij}(k_3)
 \nonumber\\&&  \times
\bigg<\phi\bigg| e^{{\rm i}(\vec k_1+\vec k_2)\cdot\vec r_1}\,\frac{1}{E-H-k_1-k_2}\,e^{{\rm i}(\vec k_3-\vec k_1)\cdot\vec r_2}\,\frac{1}{E-H-k_2-k_3}\,e^{-{\rm i}(\vec k_2+\vec k_3)\cdot\vec r_2}\nonumber\\
&&+e^{-{\rm i}(\vec k_1+\vec k_3)\cdot\vec r_2}\,\frac{1}{E-H-k_1-k_3}\,e^{-{\rm i}(\vec k_2-\vec k_3)\cdot\vec r_2}\,\frac{1}{E-H-k_1-k_2}\,e^{{\rm i}(\vec k_1+\vec k_2)\cdot\vec r_1}\nonumber\\
&&+e^{-{\rm i}(\vec k_1+\vec k_3)\cdot\vec r_2}\,\frac{1}{E-H-k_1-k_3}\,e^{{\rm i}(\vec k_1+\vec k_2)\cdot\vec r_1}\,\frac{1}{E-H-k_2-k_3}\,e^{-{\rm i}(\vec k_2-\vec k_3)\cdot\vec r_2}
 \bigg|\phi\bigg>
\,,
\end{eqnarray}
where $\delta_\perp^{ij}(k) = \delta^{ij} -k^i\,k^j/\vec k^{\,2}$.
Neglecting $E-H$ in comparison to photon energies, we express the triple seagull contribution as $E_{M1}=\langle H_{M1}\rangle$, where
\begin{align}
H_{M1}= &\ \frac{e^6\,Z^2}{m_1^{2-6\,\varepsilon}\,m_2}\int\frac{d^d k_1}{(2\pi)^d\,2k_1}
\int\frac{d^d k_2}{(2\pi)^d\,2k_2}
\int\frac{d^d k_3}{(2\pi)^d\,2k_3}\,\delta_\perp^{ik}(k_1)\delta_\perp^{ik}(k_2)\frac{(d-1)}{d}
  \nonumber\\ &\ \times e^{{\rm i}(\vec k_1+\vec k_2)\cdot\vec r}
\bigg[\frac{1}{(k_1+k_2)(k_2+k_3)} +\frac{1}{(k_1+k_3)(k_1+k_2)}+\frac{1}{(k_1+k_3)(k_2+k_3)}\bigg]\,.
\end{align}
\end{widetext}
The integration over radiative photon $k_3$ is trivial. The remaining integration is performed in spheroidal
coordinates, as explained in Appendix B of Ref.~\cite{patkos:21:rad}. The result for the triple seagull contribution is
\begin{equation}
H_{M1} = \frac{\alpha(Z\alpha)^2\pi}{2\,m_1^2\,m_2}\bigg(-\frac13+\frac43\ln2\bigg)\int \frac{d^3q}{(2\,\pi)^3}e^{i\,\vec q\cdot\vec r}\,q.
\end{equation}

\subsection{Single seagull with retardation}

The second middle-energy contribution comes from the diagram with a single seagull and
retardation, as depicted in Fig.~\ref{fig:Ei2}. Such diagram contains two photons, one of which is a transverse photon exchanged
between the electrons, and the other is a radiative photon.
The corresponding contribution to the energy is expressed as
\begin{widetext}
\begin{eqnarray}
E_{M2} &=& \frac{e^4\,Z}{m_1^{2-4\,\varepsilon}\,m_2}\int\frac{d^d k_1}{(2\pi)^d\,2k_1}\int\frac{d^d k_2}{(2\pi)^d\,2k_2}\delta_\perp^{in}(k_1)\delta_\perp^{im}(k_2)
 \nonumber\\ & \times&
\bigg<\phi\bigg|
j_1^n(k_1)\,e^{{\rm i}\vec k_1\cdot\vec r_1}\,\frac{1}{E-H-k_1}\,e^{-{\rm i}(\vec k_1+\vec k_2)\cdot\vec r_2}\,\frac{1}{E-H-k_2}\,j_2^m(k_2)\,e^{{\rm i}\vec k_2\cdot\vec r_2}\nonumber\\
&&+j_2^n(k_1)\,e^{{\rm i}\vec k_1\cdot\vec r_2}\,\frac{1}{E-H-k_1}\,e^{-{\rm i}(\vec k_1+\vec k_2)\cdot\vec r_2}\,\frac{1}{E-H-k_2}\,j_1^m(k_2)\,e^{{\rm i}\vec k_2\cdot\vec r_1}\nonumber\\
&&+j_1^n(k_1)\,e^{{\rm i}\vec k_1\cdot\vec r_1}\,\frac{1}{E-H-k_1}\,j_2^m(k_2)\,e^{{\rm i}\vec k_2\cdot\vec r_2}\,\frac{1}{E-H-k_1-k_2}\,e^{-{\rm i}(\vec k_1+\vec k_2)\cdot\vec r_2}\nonumber\\
&&+j_2^n(k_1)\,e^{{\rm i}\vec k_1\cdot\vec r_2}\,\frac{1}{E-H-k_1}\,j_1^m(k_2)\,e^{{\rm i}\vec k_2\cdot\vec r_1}\,\frac{1}{E-H-k_1-k_2}\,e^{-{\rm i}(\vec k_1+\vec k_2)\cdot\vec r_2}\nonumber\\
&&+e^{-{\rm i}(\vec k_1+\vec k_2)\cdot\vec r_2}\,\frac{1}{E-H-k_1-k_2}\,j_1^n(k_1)\,e^{{\rm i}\vec k_1\cdot\vec r_1}\,\frac{1}{E-H-k_2}\,j_2^m(k_2)\,e^{{\rm i}\vec k_2\cdot\vec r_2}\nonumber\\
&&+e^{-{\rm i}(\vec k_1+\vec k_2)\cdot\vec r_2}\,\frac{1}{E-H-k_1-k_2}\,j_2^n(k_1)\,e^{{\rm i}\vec k_1\cdot\vec r_2}\,\frac{1}{E-H-k_2}\,j_1^m(k_2)\,e^{{\rm i}\vec k_2\cdot\vec r_1}
\bigg|\phi\bigg>
\,,
\end{eqnarray}
\end{widetext}
where $j_i^l(k)$ is the current
\begin{equation}
j_i^l(k) = p^l_i + \frac{{\rm i}\,g_i}{4}\sigma_i^{kl} k^k.
\end{equation}
The $\alpha^7$ contribution is obtained by expanding the integrand up to the first order in $E-H$.
Because $ \int d^d k\,k^\alpha = 0$ in the dimensional regularization,
only the terms with $k_1+k_2$ in the denominator do not vanish, and they can be cast in the form
\begin{align}
E_{M2} = &\ -2\frac{e^4\,Z}{m_1^{2-4\,\varepsilon}\,m_2}\int\frac{d^d k_1}{(2\pi)^d\,2k_1}\int\frac{d^d k_2}{(2\pi)^d\,2k_2}\nonumber \\
&\ \times\frac{1}{k_1^2(k_1+k_2)}
\,\delta_\perp^{in}(k_1)\delta_\perp^{im}(k_2) \nonumber \\
&\ \times \bigg<\phi\bigg|
\bigg[\bigg[j_1^n(k_1)\,e^{{\rm i}\vec k_1\cdot\vec r_1},H-E\bigg],j^m_2(k_2)\,e^{{\rm i}\vec
k_2\cdot\vec r_2}\bigg] \nonumber \\
&\ \times \,e^{-{\rm i}(\vec k_1+\vec k_2)\cdot\vec r_2}
\bigg|\phi\bigg>
\,.
\end{align}
Taking into account that only the spin-independent terms survive the
double commutator and performing the angular average for the radiative photon, we arrive at
\begin{align}
E_{M2} =&\, -\frac{(4\pi\alpha)^2\,Z}{2\,m_1^{2-4\,\varepsilon}\,m_2}\frac{(d-1)}{d}\int\frac{d^d k_1}{(2\pi)^d}
\int\frac{d^d k_2}{(2\pi)^d}
 \nonumber \\ & \times \frac{1}{k_1^3k_2(k_1+k_2)}
\delta_\perp^{mn}(k_1) \langle\phi|\,e^{{\rm i} \vec k_1\cdot\vec r}\,\partial_1^m\partial_2^n
V|\phi\rangle \,.
\end{align}
We express this as the expectation value of an effective operator $H_{M2}$,
\begin{align}
H_{M2} =&\, -\frac{(4\pi\alpha)^3\,Z^2}{2\,m_1^{2-6\,\varepsilon}\,m_2}\frac{(d-1)}{d}
\int\frac{d^d q}{(2\,\pi)^d}\,e^{i\,\vec q\cdot\vec r}\int\frac{d^d k_1}{(2\pi)^d}
 \nonumber \\ & \times \int\frac{d^d k_2}{(2\pi)^d}
\frac{\delta_\perp^{mn}(k_1) q^m q^n}{k_1^3\,|\vec{k}_1-\vec{q}|^2\,k_2(k_1+k_2)} \,.
\end{align}
Performing the remaining integrations in the same way as in Ref.~\cite{patkos:21:rad}, we get the result
\begin{equation}
H_{M2} = \frac{\alpha(Z \alpha)^2\pi}{2\,m_1^{2-2\,\varepsilon}\,m_2}\int\frac{d^d q}{(2\,\pi)^d}\,e^{i\,\vec q\cdot\vec r}
\bigg(\frac49+\frac{2}{3\epsilon}-\frac83\ln \frac{q}{m_1}\bigg)\,q\,.
\end{equation}

\subsection{Total result for the middle-energy contribution}

The total result for the effective operator
representing the middle-energy contribution is
\begin{align}
H_M &\ = H_{M1} + H_{M2}
\nonumber \\ =&\
\frac{\alpha\,(Z\alpha)^2}{2\,m^{2-2\,\varepsilon}_1\,m_2}\,\pi\,
\int\frac{d^d q}{(2\,\pi)^d}\,e^{i\,\vec q\cdot\vec r} \nonumber \\
&\ \times \bigg[\,\frac19 \, +
\frac{2}{3\,\varepsilon}\,\biggl( 1 - 2\,\varepsilon\,\ln \frac{q}{2\,m_1}\biggr) - \frac43\,\ln \frac{q}{m_1} \bigg]\,q\,.
\end{align}
This needs to be transformed into the coordinate representation with the help of
\begin{align}
(\vec\nabla V)^2 =&\
-(\pi\,Z\,\alpha)^2\,m_1^{2\,\varepsilon}
\int \frac{d^d q}{(2\,\pi)^d}\,e^{i\,\vec q\cdot \vec r} \,q\,\nonumber \\
&\ \times \biggl(1-2\,\varepsilon\,\ln \frac{q}{2\,m_1}\biggr)  + O(\varepsilon^2)\,.
\end{align}
Specifically, in the limit $\varepsilon\rightarrow 0$
\begin{align}
\frac{1}{r^4} = &\ \int \frac{d^3 q}{(2\,\pi)^3}\,e^{i\,\vec q\cdot \vec r}\,(-\pi^2\,q)\,,
\\
\frac{\ln r+\gamma}{r^4} = &\  \int \frac{d^3 q}{(2\,\pi)^3}\,e^{i\,\vec q\cdot \vec r}\,\pi^2\bigg(-\frac32+\ln q\bigg)\,q\,,
\end{align}
leading to the middle-energy contribution
\begin{align}
H_M = \frac{\alpha}{2\,\pi\,m^2_1\,m_2}
\bigg[\frac{17}{9} - \frac{2}{3\,\varepsilon}-\frac43\big(\ln m_1\,r+\gamma\big)  \bigg](\vec\nabla V)^2\,,
\end{align}
where $\gamma$ is the Euler-Mascheroni constant $\gamma = 0.5772\ldots$.

\section{High-energy part}
The high-energy part $E_H$ comes from the momenta of the radiative photon of the order of electron mass $m_1$,
and is  split into three parts
\begin{align}
E_H =&\ E_{H1} + E_{H2} + E_{H3}\,,
\end{align}
where
$E_{H1}$ is due to slopes and higher derivatives of electromagnetic form factors, $E_{H2}$ is due to the anomalous magnetic moment $\kappa_1 = \alpha/(2\,\pi)$,
and $E_{H3}$ is due to QED correction to the polarizability $\alpha_E$ of the first particle beyond $\kappa_1$. 
\\

\subsection {$E_{H1}$}

The first part of the high-energy contribution comes from the derivatives $F_1'(0)$, $F_1''(0)$, and $F_2'(0)$ of electromagnetic form-factors of the first particle.
For the second particle, we assume $s_2=1/2$, and an arbitrary $g_2, r^2_{E2}, r^4_{EE2}, r^2_{M2}$, and $\alpha_{E2}$.
As a starting point we will use  Ref.~\cite{patkos:24:twobodyP} and the effective Hamiltonian
\begin{widetext}
\begin{align}
\delta H = &\ \frac{Z\alpha }{120}\,r_{EE1}^{4}\,  4\pi\,\nabla^2\delta^d(r)
	                       +\frac{Z\alpha}{24\,m_1^2} \biggl( g_1\,r_{M1}^{2} - r_{E1}^{2} \biggr)\,i\,\sigma^{ij}_1 \, p^i\, 4\pi\,\delta^d(r)\, p^j \nonumber \\
                               &\ + \frac{Z\alpha}{36}\, r_{E1}^2 \, \bigg(r_{E2}^2 + \frac{3}{4\,m_2^2}\bigg)\,4\pi\,\nabla^2\,\delta^d(r)
+i\frac{Z\alpha}{24} r_{E1}^2 \,\frac{(g_2-1)}{m_2^2}\,\sigma^{ij}_2 \, p^i\, 4\pi\,\delta^d(r)\, p^j \nonumber \\
&\ 
+ \frac{Z\alpha}{48\, m_1\,m_2}\,g_1\,r_{M1}^{2}\,\Big(2\,i\,\sigma^{ij}_1 \, p^i\, 4\pi\,\delta^d(r)\, p^j + g_2\,\sigma_1^{ik}\,\sigma_2^{jk}\,p^i\,4\pi\,\delta^d(r)\,p^j\Big)
\nonumber \\
&\ 
+ \frac{Z\,\alpha}{24\,m_1\,m_2}\,r_{E1}^2\,\bigg( 4\,\pi\, \vec\nabla^2\delta^{3}(r) + i\,g_2\,\sigma^{ij}_2 \, p^i\, 4\pi\,\delta^d(r)\, p^j \bigg) \,, 
\end{align}
\end{widetext}
where we collected all the terms that contain form-factor derivatives, given by expressions $\delta E_1-\delta E_9$ in Eqs. (36), (38), (41), (43), (45), (47)-(48), (53), and (59) of Ref.~\cite{patkos:24:twobodyP}.
The electromagnetic radii are
\begin{align}
r_{E1}^{2} = &\ \frac{6}{m_1^2} \,\Big(\frac{\kappa}{4} + F'_1(0)\Big)\,, \\
g_1\,r_{M1}^{2} = &\ \frac{12}{m_1^2} \, \big(F'_1(0) + F'_2(0)\big)\,, \\
r_{EE1}^{4} = &\ \frac{15}{m_1^4}\, \big(4\,F''_1(0) + F'_1(0) + 2\,F'_2(0)\big)\,.
\end{align}
The derivatives of form-factors are given by
\begin{align}
F'_1(0) = &\  \frac{\alpha}{\pi}\bigg[-\frac18 - \frac{1}{6\,\varepsilon}\bigg]
\,,\\
F''_1(0)= &\ \frac{\alpha}{\pi}\bigg[-\frac{11}{120} - \frac{1}{20\,\varepsilon} \bigg]
\,, \\
F'_2(0)= &\ \frac{\alpha}{12\,\pi}\,.
\end{align}
For the resulting expression $E_{H1}$ we get
\begin{widetext}
\begin{align}
E_{H1} = &\ \frac{\alpha}{\pi}\,\frac{Z\alpha}{m_1^2}\bigg\langle \bigg\{-\frac{1}{m_1^2}\bigg[\frac{13}{160} + \frac{11}{120\,\varepsilon}\bigg]
 - \frac{1}{m_2^2}\bigg[\frac{1}{32} +\frac{1}{24\,\varepsilon} \bigg] - \frac{1}{m_1\,m_2}\bigg[\frac{1}{16} + \frac{1}{12\,\varepsilon} \bigg]
-\frac{g_2\,\sigma_1^{ij}\sigma_2^{ij}}{m_1\,m_2}\bigg[\frac{11}{864} + \frac{1}{72\,\varepsilon}\bigg] \nonumber \\
&\    - r_{E2}^2\,\bigg[\frac{1}{24} + \frac{1}{18\,\varepsilon}\bigg]\bigg\}\,\vec p\,4\pi\,\delta^d(r)\,\vec p
-\frac{g_2}{m_1\,m_2}\bigg[\frac{1}{96} + \frac{1}{24\,\varepsilon}\bigg]\sigma_1^{ik}\,\sigma_2^{jk}\,\big(p^i\,4\pi\,\delta^d(r)\,p^j\big)^{(2)} \nonumber \\
&\ - {\rm i}\,\bigg\{-\frac{\sigma_1^{ij}}{m_1^2}\bigg[\frac{1}{96}-\frac{1}{24\,\varepsilon}\bigg] + 
\frac{(g_2-1)\,\sigma_2^{ij}}{2\,m_2^2}\bigg[\frac{1}{32} + \frac{1}{24\,\varepsilon}\bigg] 
+ \frac{\sigma_1^{ij}}{m_1\,m_2}\bigg[\frac{1}{48} + \frac{1}{12\,\varepsilon}\bigg] \bigg\} \,
p^i\,4\pi\,\delta^d(r)\,p^j\bigg\rangle\,.\label{43}
\end{align}
\end{widetext}
For the case of two pointlike particles, we checked this result also by a complementary method of calculation, namely the scattering amplitude approach, 
as was done for the $E_9$ contribution in Ref.~\cite{patkos:21:rad}.
Generalizing the derivation in Ref.~\cite{patkos:21:rad} for arbitrary masses of both particles and considering also the spin-orbit terms, we get the result in agreement with Eq.~(\ref{43})
for the pointlike second particle.

\subsection{$E_{H2}$}
$E_{H2}$ is the contribution due to the anomalous magnetic moment (amm) $\kappa$ of the pointlike first particle. It can be obtained by collecting all the $\kappa$-dependent
parts of the first-order operators $\delta E_1 - \delta E_9$ in Ref.~\cite{patkos:24:twobodyP}, where $\kappa$ is present in the $g$ factor $g=2\,(1+\kappa)$ and in 
the electric dipole polarizability  \begin{align}
\delta H = -\frac{e^2}{2}\,\alpha_E\,\vec E^{\,2}
\end{align}
where
\begin{equation}
\alpha_{E} = -\frac{\kappa\,(1+\kappa)}{4\,m^3} -\frac{\alpha}{3\,\pi\,m^3}\biggl(1-\frac{2}{\varepsilon}\biggr)\,. \label{epol}
\end{equation}
We shall add a few comments at this point. If we consider a point particle with the magnetic moment anomaly, then 
the electric dipole polarizability includes the first term in the above equation. 
The additional radiative correction, which is not accounted for by the magnetic moment anomaly, 
is the second term, which was calculated in Ref.~\cite{jentschura:05}.
Here, we account only for the first term, and in the next subsection, we will separately address the second term.
This is because, for a non-point particle such as an antiproton, we will include the first term in the definition of the electric dipole polarizability,
and the second term will be an additional correction with $1/\varepsilon$ infrared singularity to be canceled with a similar term in the low-energy part.

All these contributions due to the magnetic moments are finite, and thus we may present them in three-dimensional form as
\begin{align}\label{amm:total}
E_{H2} = &\ \kappa_1\,\biggl(\sum_{i=1\ldots 9} \langle \delta H_i\rangle + E_{\rm sec}\biggr)\,,
\end{align}
where the individual $\delta H_i$ operators were derived in Ref.~\cite{patkos:24:twobodyP} and are presented in Appendix \ref{app:ops}.
$E_\mathrm{sec}$ is a second-order amm contribution
\begin{align}\label{amm:sec}
E_{\rm sec} = &\ 2\,\bigg\langle H^{(4)}_{\rm amm}\,\frac{1}{(E-H)'}\,H^{(4)}(\kappa_1=0)\bigg\rangle\nonumber \\
= &\ \frac{\alpha}{\pi}\bigg[ E_{\rm sec}^\mathrm{NS} + E_{\rm sec}^\mathrm{S1}\,\langle\vec L\cdot\vec s_1\rangle 
+ E_{\rm sec}^\mathrm{S2}\,\langle\vec L\cdot\vec s_2\rangle \nonumber \\
&\ + E_{\rm sec}^\mathrm{SS}\,\langle\vec s_1\cdot\vec s_2\rangle + E_{\rm sec}^\mathrm{LL}\,\langle(L^i\,L^j)^{(2)}\,s_1^i\,s_2^j\rangle\bigg]
\end{align}
where $H^{(4)}_{\rm amm}$ is the part of $H^{(4)}$ in Eq.~(\ref{Breitgeneral}) which is linear in $\kappa_1$,
and $H^{(4)}(\kappa_1=0)$ is the Breit Hamiltonian with $\kappa_1$ omitted.

\subsection{$E_{H3}$}
This is a correction due to the second term in the electric dipole polarizability in Eq. (\ref{epol}),
\begin{align}
E_{H3} = \frac{\alpha}{\pi\,m_1^3}\bigg(\frac16 - \frac{1}{3\,\varepsilon}\bigg)(\vec\nabla V)^2\,,
\end{align}
which is considered separately because it is infrared divergent. We will assume that it is common to all particles, including
all nuclei, and will exclude it from the definition of the electric dipole polarizability.

\section{Total one-loop radiative correction}
With the help of the identity derived in Appendix \ref{app:identities} valid for $l>0$ states,
\begin{align}\label{83}
&\ p^i\,\bigg[\frac{Z\alpha}{r^3}\bigg(\delta^{ij} - 3\frac{r^i \,r^j}{r^2}\bigg)\bigg]_\epsilon\,p^j \nonumber \\
=&\ Z\alpha\bigg(\frac16 - \frac{2}{9}\,\varepsilon\bigg)\,\vec p\,4\pi\,\delta^d(r)\,\vec p + \mu\,(\vec \nabla V)^2\,,
\end{align}
all the singularities proportional to $1/\varepsilon$ cancel out  algebraically in the sum of all parts in Eq. (\ref{03}). We may 
therefore pass to three dimensions by setting $\varepsilon\rightarrow 0$ and replace
\begin{align}
\vec p\,4\pi\,\delta^d(r)\,\vec p \rightarrow &\ \vec p\,4\pi\,\delta^3(r)\,\vec p\,,\\
\sigma_a^{ij}\,p^i\,4\pi\,\delta^d(r)\,p^j \rightarrow &\ 2\,\vec s_a\cdot \vec p\times 4\pi\,\delta^3(r)\,\vec p\,,\\
\sigma_1^{ik}\,\sigma_2^{jk}\,p^i\,4\pi\,\delta^d(r)\,p^j \rightarrow &\ 4\,\vec{s}_2 \times \vec{p} \,4\pi\,\delta^3(r)\,\vec{s}_1 \times \vec{p}\,.
\end{align}
The final expression for $E^{(7)}_\mathrm{rad}$ in Eq. (\ref{03})
for the $\alpha^7$ radiative two-body correction to the energy is
\begin{align}\label{E7rad ops}
E^{(7)}_\mathrm{rad} = &\ \frac{\alpha}{\pi}\,\big\langle E_\mathrm{NS} + \vec L\cdot\vec s_1\, E_\mathrm{S1}
+\vec L\cdot\vec s_2 \, E_\mathrm{S2} + \vec s_1\cdot\vec s_2 \, E_\mathrm{SS} \nonumber \\
&\ + (L^i\,L^j)^{(2)}\,s_1^i\,s_2^j \,E_\mathrm{LL} \big\rangle\,,
\end{align}
where individual coefficients  are
\begin{widetext}
\begin{align}
E_\mathrm{NS} = &\
\bigg\langle 
\frac{Z\alpha}{m_1^4\,m_2^2}\bigg[\bigg(\frac{31}{288} + \frac16\,\LZasquaredmmu \bigg)\,m_1\,m_2
+\bigg(\frac{779}{7200}+\frac{11}{60}\,\LZasquaredmmu\bigg)\,m_2^2 \nonumber\\
&\ + \frac{1}{72}\,\bigg(1 + \frac43\,m_2^2\,r_{E2}^2\bigg)\,\big(5+6\,\LZasquaredmmu \big)\,m_1^2\bigg]\,\vec p\,4\pi\,\delta^3(r)\,\vec p
+ \frac{1}{m_1^3\,m_2\,(m_1+m_2)} \nonumber \\
&\ \times \bigg[\bigg(\frac{17}{12} + \frac23\,\LZasquaredmmu - \frac23\,\big(\ln m_1\, r+\gamma\big)\bigg)\,m_1^2
+\bigg(\frac{589}{720}+\frac23\,\LZasquaredmmu\bigg)\,m_2^2\nonumber \\
&\ +m_1\,m_2\,\bigg(\frac{317}{144}+\frac43\,\LZasquaredmmu
-\frac23\,\big(\ln m_1\, r+\gamma\big) \bigg)\bigg]\,\frac{(Z\alpha)^2}{r^4} 
 \bigg\rangle + E_{\rm sec}^\mathrm{NS} + \frac{(Z\alpha)^6\,\mu}{n^3}\,\beta^\mathrm{NS}\,,\\
E_\mathrm{S1} = &\ \bigg\langle- \frac{(m_1^2 - m_1\,m_2 + m_2^2)}{2\,m_1^3\,m_2^2}\frac{E\,Z\alpha}{r^3}
- \frac{(2\,m_1^2 + m_1\,m_2 + 2\,m_2^2)}{4\,m_1^3\,m_2^2}\frac{(Z\alpha)^2}{r^4}
 \nonumber \\
&\ - \frac{Z\alpha}{m_1^4\,m_2^3}\,\bigg[ \bigg(\frac{43}{144} + \frac13\,\LZasquaredmmu\bigg)\,m_1\,m_2^2 
+ \bigg(\frac{23}{144} + \frac16\,\LZasquaredmmu\bigg)
\,m_2^3 \nonumber \\
&\ + \frac{1}{16}\,m_1^2\,m_2 + \frac{1}{32}\,m_1^3\,(2 - g_2 ) + \frac{m_1^3\,m_2^3}{12\,\mu}\,r_{E2}^2 \bigg]\,\vec p\,4\pi\,\delta^3(r)\,\vec p 
\bigg\rangle + E_{\rm sec}^\mathrm{S1} + \frac{(Z\alpha)^6\,\mu}{n^3}\,\beta^\mathrm{S1}\,,\\
E_\mathrm{S2} = &\  \bigg\langle- \frac{Z\alpha}{m_1^3\,m_2^2} \bigg[\bigg(\frac{5}{36} + \frac16\,\LZasquaredmmu\bigg)
(g_2-1)\,m_1 + \bigg(\frac{31}{288} + \frac16\,\LZasquaredmmu\bigg)\,g_2\,m_2\bigg]\,\vec p\,4\pi\,\delta^3(r)\,\vec p \bigg\rangle \nonumber \\
&\ + E_\mathrm{sec}^\mathrm{S2} + \frac{(Z\alpha)^6\,\mu}{n^3}\,\beta^\mathrm{S2}\,,\\
E_\mathrm{SS} = &\ \bigg\langle\frac{\big((5\,g_2-6)\,m_1 + 5\,g_2\,m_2\big)}{24\,m_1^2\,m_2^2}\frac{(Z\alpha)^2}{r^4} 
 + \frac{Z\alpha}{m_1^3\,m_2^3}\bigg[\frac{(g_2-2)}{48}\,m_1^2 + \frac{(g_2-1)}{24}\,m_1\,m_2 
 \nonumber \\
&\
 + \bigg(\frac{77}{432} + \frac{2}{9}\,\LZasquaredmmu\bigg)  \,g_2\,m_2^2 + \frac{g_2}{18}\,m_1^2\,m_2^2\,r_{M2}^2\bigg]\,\vec p\,4\pi\,\delta^3(r)\,\vec p
 \bigg\rangle +  E_{\rm sec}^\mathrm{SS} + \frac{(Z\alpha)^6\,\mu}{n^3}\,\beta^\mathrm{SS} \,,\\
E_\mathrm{LL} = &\ \frac{Z\alpha}{(2l-1)(2l+3)}\,\bigg\langle \frac{3\,(m_1 + m_2 - g_2\,m_2)}{m_1\,m_2^2\,(m_1+m_2)}\frac{E}{r^3}
+\frac{\big[(12+g_2)\,m_1^2 + (12-7\,g_2)\,m_1\,m_2 + g_2\,m_2^2\big]}{4\,m_1^2\,m_2^2\,(m_1+m_2)}\frac{Z\alpha}{r^4} \nonumber \\
&\ + \frac{1}{m_1^3\,m_2^3}\bigg[\frac{7\,(g_2-2)}{16}\,m_1^2 + \frac{5\,(g_2-1)}{4}\,m_1\,m_2 + \bigg(\frac{233}{144}
+\frac53\,\LZasquaredmmu\bigg)\,g_2\,m_2^2
+\frac{5\,g_2}{12}\,m_1^2\,m_2^2\,r_{M2}^2\bigg]
\nonumber \\
&\ \times \vec p\,4\pi\,\delta^3(r)\,\vec p \bigg\rangle  + E_{\rm sec}^\mathrm{LL} + \frac{(Z\alpha)^6\,\mu}{n^3}\,\beta^\mathrm{LL}\,.
\end{align}
\end{widetext}
The expectation values of the first-order operators are evaluated with the help of formulas from Appendix \ref{app:formulas}.
The second-order contribution, which comes exclusively from the amm contribution, is evaluated in the same way as in Ref.~\cite{zatorski:22}.
We will now present the final formula for the radiative $\alpha^7$ contribution to energy.

\section{Results}

The general result can be cast in the form

\begin{align}\label{res}
E^{(7)}_\mathrm{rad} = &\ \frac{\mu\,\alpha(Z\alpha)^6}{\pi\,\Delta}\big\langle \mathcal{E}_\mathrm{NS} + \vec L\cdot\vec s_1\,\mathcal{E}_\mathrm{S1}
+ \vec L\cdot\vec s_2\,\mathcal{E}_\mathrm{S2} \nonumber \\
&\ + \vec s_1\cdot\vec s_2\,\mathcal{E}_\mathrm{SS}
+ (L^i L^j)^{(2)}\,s_1^i s_2^j\,\mathcal{E}_\mathrm{LL}
\big\rangle  \,,
\end{align}
where we pulled out the factor $\Delta^{-1}$, with $\Delta = 30$ for $l=1$, and for $l>1$ it is defined in Eq.~(\ref{delta}).
We consider separately the cases with $l=1$ and $l>1$, where for the latter case
the individual coefficients are lengthy and thus we move their explicit results into Appendix \ref{app:results}.
Defining
\begin{equation}
\eta_1 = \frac{\mu}{m_1},\,\,\,\eta_2 = \frac{\mu}{m_2}\,,\,\,\ln_1 = \LZasquaredeta\,
\end{equation}
the results for $l=1$ are
\begin{align}
\mathcal{E}_\mathrm{NS} = &\
\frac{\mathcal{E}_\mathrm{NS}^{(3)}}{n^3} + \frac{\mathcal{E}_\mathrm{NS}^{(4)}}{n^4} + \frac{\mathcal{E}_\mathrm{NS}^{(5)}}{n^5}
+ 8\,\eta_1^2\,\eta_2\,\bigg(\frac{2}{3\,n^5} - \frac{1}{n^3}\bigg) \nonumber \\
&\ \times\bigg( \frac{137}{60} - H_{n+1} +\ln\frac{n}{2\,\eta_1\,Z\alpha}\bigg)
+ \frac{100\,\mu^2\,\eta_1^2}{27}\,r_{E2}^2\, \nonumber \\
&\ \times \bigg(\frac{1}{n^3}-\frac{1}{n^5}\bigg)+ \ln_1\,\mathcal{E}_\mathrm{NS}^\mathrm{log} + \frac{\Delta}{n^3}\,\beta_\mathrm{NS}\,, \label{81}
\\
\mathcal{E}_\mathrm{NS}^{(3)} = &\ \eta_1^2\bigg(\frac{821}{36} - \eta_1 \frac{1081}{72}+\eta_1^2 \frac{16}{5}\bigg)  
-\eta_1^2\,\eta_2^2 \frac{137}{480} g_2^2\,,\\
\mathcal{E}_\mathrm{NS}^{(4)} = &\ \eta_1^2\bigg(-\frac{29}{2} + \eta_1\,\frac{53}{4}\bigg)  - \eta_1^2\,\eta_2^2\,\frac{3 }{16}\,g_2^2\,, \\
\mathcal{E}_\mathrm{NS}^{(5)} = &\ \eta_1^2\bigg(-\frac{112}{9} + \eta_1\,\frac{221}{36} - \eta_1^2\,\frac{46}{15}\bigg)\,, \\
\mathcal{E}_\mathrm{NS}^\mathrm{log} = &\ \eta_1^2\bigg[\bigg(\frac{1}{n^3} - \frac{1}{n^5}\bigg)
\bigg(  \frac{34}{3} + 4 \eta_1^2
 + \frac{40 \mu^2}{9}r_{E2}^2\bigg)
 + \frac{8}{3\,n^5}\bigg]\,, \\
\mathcal{E}_\mathrm{S1} = &\
\frac{\mathcal{E}_\mathrm{S1}^{(3)}}{n^3} + \frac{\mathcal{E}_\mathrm{S1}^{(4)}}{n^4} + \frac{\mathcal{E}_\mathrm{S1}^{(5)}}{n^5} 
- \bigg(\frac{1}{n^3} - \frac{1}{n^5}\bigg)\bigg(\frac{10\,\mu^2\,\eta_1}{3}\,r_{E2}^2 \nonumber \\
&\ +\frac{20 \,\eta_1^3(\eta_2 + 1)}{3} \ln_1 \bigg)  + \frac{\Delta}{n^3}\beta_\mathrm{S1}\,, \\
\mathcal{E}_\mathrm{S1}^{(3)} = &\ \eta_1\bigg(\frac{10}{9} + \eta_1\,\frac{191}{36} - \eta_1^2\,\frac{835}{72} + \eta_1^3\,\frac{50}{9}\bigg)  + \eta_1\eta_2^3\frac{133}{288}g_2  \nonumber \\
&\   
+g_2^2\,\eta_1\,\eta_2^2\bigg(\frac{227}{288} -\eta_1\,\frac{13}{320} \bigg) \,,\\
\mathcal{E}_\mathrm{S1}^{(4)} = &\  
  - \eta_1\,\eta_2^3 \frac{5}{16}\,g_2 + g_2^2\,\eta_1\,\eta_2^2\bigg(\frac{5}{16} + \eta_1\, \frac{9}{32}\bigg) \nonumber \\
  &\  + \eta_1 \bigg(\frac{25}{4} - \eta_2^2\frac54\bigg)\,,\\
\mathcal{E}_\mathrm{S1}^{(5)} = &\ \eta_1\bigg(-\frac{15}{2} + \eta_1 + \eta_1^2\,\frac{125}{18} - \eta_1^3\,\frac{50}{9} \bigg) 
\nonumber \\ &\ 
 - \eta_1\,\eta_2^3 \bigg( \frac78 g_2 + \frac38 g_2^2 \bigg)\,,  \\
\mathcal{E}_\mathrm{S2} = &\
\frac{\mathcal{E}_\mathrm{S2}^{(3)}}{n^3} + \frac{\mathcal{E}_\mathrm{S2}^{(4)}}{n^4} + \frac{\mathcal{E}_\mathrm{S2}^{(5)}}{n^5}
+ \frac{\Delta}{n^3}\,\beta_\mathrm{S2}
-\frac{20\,\eta_1^2\,\eta_2}{3}\big(g_2 - \eta_2\big)\nonumber \\
&\ \times \bigg(\frac{1}{n^3} - \frac{1}{n^5}\bigg)\,\ln_1 \,, \\
\mathcal{E}_\mathrm{S2}^{(3)} = &\ \eta_1^2\,\eta_2^2 \bigg(\frac{50}{9} - \frac{13}{320}g_2^2\bigg) 
-g_2\,\eta_1^2\,\eta_2\bigg(\frac{559}{288} + \eta_2\,\frac{133}{288}\bigg) 
 \,, \\
\mathcal{E}_\mathrm{S2}^{(4)} = &\ g_2\,\eta_1^2\,\eta_2\bigg(\frac{15}{16}+\eta_2\,\frac{5}{16} \bigg)  + g_2^2\,\eta_1^2\,\eta_2^2\frac{9}{32}, \\
\mathcal{E}_\mathrm{S2}^{(5)} = &\ g_2\,\eta_1^2\,\eta_2\bigg( \frac{229}{72} + \eta_2\,\frac78 \bigg) 
 - \eta_1^2\,\eta_2^2\,\bigg(\frac{50}{9} - \frac38\,g_2^2\bigg)\,,\\
\mathcal{E}_\mathrm{SS} = &\
\frac{\mathcal{E}_\mathrm{SS}^{(3)}}{n^3} + \frac{\mathcal{E}_\mathrm{SS}^{(4)}}{n^4} + \frac{\mathcal{E}_\mathrm{SS}^{(5)}}{n^5} 
+ \frac{\Delta}{n^3}\,\beta_\mathrm{SS}
+ \frac{20\,g_2}{9}\bigg(\frac{1}{n^3} - \frac{1}{n^5}\bigg) \nonumber \\
&\ \times \bigg(\mu^2\,\eta_1\,\eta_2
\,r_{M2}^2+4\,\eta_1^3\,\eta_2\,\ln_1 \bigg) \,, \\
\mathcal{E}_\mathrm{SS}^{(3)} = & \ g_2\,\eta_1\,\eta_2\bigg(-\frac{47}{54} +\eta_1^2 \frac{170}{27}\bigg) 
 - \eta_1^2\,\eta_2^2\,\frac{137}{360}\,g_2^2 \nonumber \\ &\  - \eta_1\eta_2^2\frac{25}{54} \,, \\
\mathcal{E}_\mathrm{SS}^{(4)} = & \ - \eta_1\,\eta_2\,\frac{5 }{3}\,g_2 + \eta_1\,\eta_2^2\,\frac53  
   - \eta_1^2\,\eta_2^2\,\frac{g_2^2}{4}\,, \\
\mathcal{E}_\mathrm{SS}^{(5)} = & \ g_2\,\eta_1\,\eta_2\bigg(-\frac12 -  \eta_1^2\,\frac{170}{27}\bigg)   + \eta_1\,\eta_2^2\,\frac53
\,, \\
\mathcal{E}_\mathrm{LL} = &\
\frac{\mathcal{E}_\mathrm{LL}^{(3)}}{n^3} + \frac{\mathcal{E}_\mathrm{LL}^{(4)}}{n^4} + \frac{\mathcal{E}_\mathrm{LL}^{(5)}}{n^5} 
+ \frac{\Delta}{n^3}\,\beta_\mathrm{LL} + \frac{10\,g_2}{3}\bigg(\frac{1}{n^3} - \frac{1}{n^5}\bigg) \nonumber \\
&\ \times\bigg(\mu^2\,\eta_1\,\eta_2\,r_{M2}^2 +4\,\eta_1^3\,\eta_2\,\ln_1 \bigg) \,, \\
\mathcal{E}_\mathrm{LL}^{(3)} = &\ \eta_1\,\eta_2^2\bigg(\frac{1171}{180}  - 3\eta_1\bigg) -
g_2\,\eta_1\,\eta_2\bigg(\frac{3697}{720}  + \eta_1\frac{417}{40}\nonumber \\
&\
- \eta_1^2\frac{3067}{360}\bigg) +g_2^2\,\eta_1\,\eta_2^2\bigg(-\frac{227}{80}+\eta_1 \frac{1291}{1200}\bigg)   \,, \\
\mathcal{E}_\mathrm{LL}^{(4)} = &\  \eta_1\eta_2^2\frac52 + g_2\,\eta_1\,\eta_2\bigg(-\frac{71}{8} + \eta_2\frac{9}{4} + \eta_2^2\frac94\bigg)
\nonumber \\
&\ + g_2^2\,\eta_1\,\eta_2^2\bigg( -\frac98 -\eta_1\,\frac{21}{40}\bigg) \,, \\
\mathcal{E}_\mathrm{LL}^{(5)} = &\ \eta_1\,\eta_2^2\bigg(-\frac{19}{5} + 3\,\eta_1\bigg)
+ g_2\,\eta_1\,\eta_2\bigg( \frac{153}{20} - \eta_1\,\frac{9}{5} \nonumber \\
&\ - \eta_1^2\,\frac{101}{45}\bigg)
 + \eta_1\,\eta_2^2\,(4\,\eta_2 - 1)\,\frac{9}{20}\,g_2^2\,, \label{101}
\end{align}
where $H_n = \sum_{i=1}^n i^{-1}$ gives the $n$-th harmonic number.
The Bethe logarithmic terms will be calculated after combining them with those from 
the exchange contribution at $(Z\,\alpha)^7$ order.
We now consider special cases of the general results in Eq.~(\ref{res}).

\subsection{Positronium}

First, we will examine the case of a positronium atom, i.e., the two-body system of bound electron and positron.
To achieve this, we treat the nucleus as pointlike by setting $g_2=2$, $r_{E2}^2 = r_{M2}^2= 0$, $m_1 = m_2 = m$, and
include the corresponding result for the radiative correction of the second particle, where we make the exchange $(1\leftrightarrow2)$.
For the $l=1$ states we get the result
\begin{align}
E^{(7)}_\mathrm{pos}(n^S P_J) = &\ \frac{\alpha(Z\alpha)^6\,m}{\pi}\bigg[\mathcal{ E}^{(7)}(n^S P_J) 
\nonumber \\
&\ + \bigg(\frac{1}{30\,n^3} - \frac{1}{45\,n^5}\bigg)\bigg(H_{n+1} - \ln\frac{n}{ Z\alpha}\bigg) \bigg] \,, \label{102}
\end{align}
where
\begin{align}
 \mathcal{ E}^{(7)}(n^1P_1) = &\ \frac{\beta^\mathrm{pos}({}^1P_1)}{n^3} + \frac{73}{1440\,n^3}
-\frac{1}{20\,n^4} - \frac{47}{3600\,n^5}  \nonumber \\
&\ + \bigg(\frac{3}{40\,n^3} - \frac{19}{360\,n^5}\bigg)\LZasquaredpos \,, \\
 \mathcal{ E}^{(7)}(n^3P_0) = &\ \frac{\beta^\mathrm{pos}({}^3P_0)}{n^3} -\frac{101}{960\,n^3}
-\frac{3}{10 n^4} + \frac{307}{2400\,n^5}\nonumber \\
&\    +  \bigg( \frac{29}{120\,n^3} - \frac{79}{360\,n^5}\bigg) \LZasquaredpos \,, \\
 \mathcal{ E}^{(7)}(n^3P_1) = &\ \frac{\beta^\mathrm{pos}({}^3P_1)}{n^3} + \frac{181}{1728 n^3} 
-\frac{73}{960 n^4} - \frac{877}{21600 n^5} \nonumber \\
&\   + \bigg(\frac{47}{360\,n^3} - \frac{13}{120\,n^5}\bigg)\LZasquaredpos \,, \\
 \mathcal{ E}^{(7)}(n^3P_2) = &\ \frac{\beta^\mathrm{pos}({}^3P_2)}{n^3} + \frac{491}{8000\,n^3} 
-\frac{41}{1600\,n^4} - \frac{67}{900\,n^5} \nonumber \\
&\    + \bigg(\frac{3}{40\,n^3} - \frac{19}{360\,n^5}\bigg)\LZasquaredpos  \,, \\
 \beta^\mathrm{pos}({}^{2s+1}P_j) = &\ \beta_\mathrm{NS}(1) + F\,\big(\beta_\mathrm{S1}(1) + \beta_\mathrm{S2}(1)\big)  \nonumber \\
&\ +  \frac{\big( 2\,s(s+1) -3\big)}{4}\,\beta_{SS}(1) \nonumber \\
&\ +\big[ 3\,F\,(1+2\,F) - 4\,s(s+1) \big]\,\frac{\beta_\mathrm{LL}(1)}{12}\,, \\
F = &\ \frac12\,\big[j(j+1) - s(s+1) - 2\big]\,, \label{108}
\end{align}
where we introduced notation $\beta_i(x) = \beta_i$ with $x=m_1/m_2$. 

\subsection{Hydrogenlike atoms}

For hydrogenlike atoms, we begin with the nonrecoil limit, assuming the nuclear mass $m_2$ to be infinitely heavy. 
We consider the case of $l=1$ while the $l>1$ case is presented in Appendix \ref{app:results}. 
We obtain the result

\begin{align}
E^{(7,0)}_{\rm hydr}(n P) = &\ \frac{m_1\,\alpha(Z\alpha)^6}{\pi} 
\big\langle \mathcal{E}_\mathrm{NS}^{(7,0)} 
+ \vec L\cdot\vec s_1\,\mathcal{E}_\mathrm{S1}^{(7,0)} \big\rangle\,, \\
\mathcal{E}_\mathrm{NS}^{(7,0)} = &\  \frac{1319}{3600\,n^3} 
  -\frac{1}{24\,n^4} - \frac{1687}{5400\,n^5} \nonumber \\ &\
  + \frac{10}{81} \bigg(\frac{1}{n^3}-\frac{1}{n^5}\bigg)\,m_1^2\,r_{E2}^2
  +\frac{\beta_\mathrm{NS}^{(0)}}{n^3} \nonumber \\
&\ + \LZasquared\bigg[\frac{23}{45\,n^3} - \frac{19}{45\,n^5} \nonumber \\ &\
+ \frac{4}{27}\,m_1^2\,r_{E2}^2\bigg(\frac{1}{n^3}-\frac{1}{n^5}\bigg)\bigg]\,,\\
\mathcal{E}_\mathrm{S1}^{(7,0)} = &\  \frac{1}{80\,n^3} 
+\frac{5}{24\,n^4} -\frac{23}{135\,n^5} \nonumber \\ &\
-  \frac19 \bigg(\frac{1}{n^3}-\frac{1}{n^5}\bigg)\,m_1^2\,r_{E2}^2
  +\frac{\beta_\mathrm{S1}^{(0)}}{n^3} \nonumber \\
&\ - \frac29\,\bigg(\frac{1}{n^3}-\frac{1}{n^5}\bigg)\,\LZasquared \,,
\end{align}
where
\begin{equation}
\beta_i(x) = \beta_i^{(0)} + x\,\beta_i^{(1)} + x^2\,\beta_i^{(2)} + \ldots\,.
\end{equation}
This result is in agreement with the one from Ref.~\cite{jentschura:05} for a pointlike nucleus.
For the leading recoil contribution we get
\begin{align}\label{hydrrec}
E^{(7,1)}_{\rm hydr}(n P) = &\
\frac{m_1^2 \alpha(Z\alpha)^6}{\pi\,m_2}\big\langle \mathcal{E}_\mathrm{NS}^{(7,1)} + \vec L\cdot\vec s_1 \mathcal{E}_\mathrm{S1}^{(7,1)}
+\vec L\cdot\vec s_2 \mathcal{E}_\mathrm{S2}^{(7,1)} \nonumber \\ &\
+ \vec s_1\cdot\vec s_2\,\mathcal{E}_\mathrm{SS}^{(7,1)}
+(L^i L^j)^{(2)}\,s_1^i \,s_2^j\,\mathcal{E}_\mathrm{LL}^{(7,1)} \big\rangle\,, \\
\mathcal{E}_\mathrm{NS}^{(7,1)} = &\ \frac{\beta^{(1)}_\mathrm{NS} - \beta^{(0)}_\mathrm{NS}}{n^3} -\frac{4913}{5400\,n^3}
-\frac{19}{60\,n^4} + \frac{1243}{1350\,n^5}  \nonumber \\ &\
+ \bigg(\frac{4}{15\,n^3} - \frac{8}{45\,n^5}\bigg)\bigg(H_{n+1} - \ln\frac{n}{2\, Z\alpha}\bigg) \nonumber \\ &\
-\frac{38}{81} \bigg(\frac{1}{n^3} - \frac{1}{n^5}\bigg)\,m_1^2\,r_{E2}^2 + \bigg[  \frac{23}{15\,n^5}-\frac{9}{5\,n^3} \nonumber \\
&\ -\frac{20}{27} m_1^2\,r_{E2}^2 \bigg(\frac{1}{n^3} - \frac{1}{n^5}\bigg)\bigg]\,\LZasquared \,, \\
\mathcal{E}_\mathrm{S1}^{(7,1)} = &\ \frac{\beta^{(1)}_\mathrm{S1} - \beta^{(0)}_\mathrm{S1}}{n^3} -\frac{223}{1080\,n^3} - \frac{5}{12\,n^4} + \frac{28}{45\,n^5} \nonumber \\ &\
+\frac49\,\bigg(\frac{1}{n^3} - \frac{1}{n^5}\bigg)\,m_1^2\,r_{E2}^2 + \frac23\,\bigg(\frac{1}{n^3} - \frac{1}{n^5}\bigg)\nonumber \\
&\ \times\LZasquared \,,\\
\mathcal{E}_\mathrm{S2}^{(7,1)} = & \frac{\beta^{(1)}_\mathrm{S2} }{n^3} + g_2\bigg(-\frac{559}{8640\,n^3} + \frac{1}{32\,n^4} + \frac{229}{2160\,n^5}\bigg) \nonumber \\
&\ - g_2\,\frac29\,\bigg(\frac{1}{n^3} - \frac{1}{n^5}\bigg)\,\LZasquared
 \,,\\
\mathcal{E}_\mathrm{SS}^{(7,1)} = &\ \frac{\beta^{(1)}_\mathrm{SS} }{n^3} + g_2\bigg(\frac{293}{1620\,n^3} - \frac{1}{18\,n^4} - \frac{367}{1620\,n^5}\bigg) \nonumber \\ &\
 +g_2\,\frac{2}{27}\,\bigg(\frac{1}{n^3} - \frac{1}{n^5}\bigg)\,m_1^2\,r_{M2}^2 \nonumber \\
&\ + g_2\,\frac{8}{27}\,\bigg(\frac{1}{n^3} - \frac{1}{n^5}\bigg)\,\LZasquared\,,\\
 \mathcal{E}_\mathrm{LL}^{(7,1)} = &\ \frac{\beta^{(1)}_\mathrm{LL} }{n^3} +  g_2\bigg(-\frac{5069}{21\,600\,n^3} - \frac{71}{240\,n^4}  \nonumber \\ &\
 + \frac{649}{5400\,n^5}\bigg) +g_2\,\frac{1}{9}\,\bigg(\frac{1}{n^3} - \frac{1}{n^5}\bigg)\,m_1^2\,r_{M2}^2 \nonumber \\
 &\ + g_2\,\frac49\,\bigg(\frac{1}{n^3} - \frac{1}{n^5}\bigg)\,\LZasquared\,,
\end{align}

\section{Antiprotonic atoms}

We may apply the results of our calculation also to  highly excited rotational states of antiprotonic atoms.
In the case of a two-body system consisting of two hadronic particles, one has to include the strong interaction
effects. However, for highly excited rotational states, these  effects are negligible due to their short range.
We may also omit all the other  local interaction terms, but we have to keep the $g$ factor of the first particle in the general form,
and include also the radiative contribution for the second (heavy) particle. As a result,
only the low-energy, middle-energy, and $E_{H3}$ contributions have to be taken into account.

For antiprotonic atoms, the low-energy contribution $E_{L1}$ is 
\begin{align}
\label{EL1intermediateAnt}
&\ E_{L1} =
\frac{\alpha}{3\,\pi\,m_1^2}\,
\left\{\frac{1}{2\,\varepsilon}+\frac{5}{6} +
\LZasquaredmmu\right\}\nonumber \\
&\ \times \bigg< 
-\frac{Z\alpha}{m_1\,m_2}\,p^i\,\bigg(\frac{\delta^{ij}}{r^3} - 3\frac{r^i \,r^j}{r^5}\bigg)_\epsilon\,p^j
\bigg> \nonumber \\
&\ + \frac{\alpha}{\pi}\,\frac{(Z\,\alpha)^6}{n^3}\,\mu\,\beta_1(x) +Z^2(1\leftrightarrow2,x\leftrightarrow x^{-1})\,.
\end{align}
In the Bethe logarithm  contribution the perturbation of the
expectation value by the Breit Hamiltonian $H^{(4)}$ has to include $g$-factors of both particles.

The low-energy contribution $E_{L2}$ is for antiprotonic atoms of the form
\begin{align}
\label{El2intermediateAnt}
E_{L2}  = &\
\frac{\alpha}{\pi}\,\bigg<
(\vec{\nabla} V)^2\,
\bigg[\frac{1}{m_1^3}\bigg(\frac{1}{2\varepsilon} +\frac{5}{6}  +\LZasquaredmmu\bigg)\nonumber \\
&\   + \frac{\mu}{m_1^4}
\bigg(\frac{1}{6\varepsilon} +\frac{14}{45}+\frac13
\LZasquaredmmu\bigg)\bigg]
\bigg>\nonumber \\
&\ +\frac{\alpha}{\pi}\,\frac{(Z\,\alpha)^6}{n^3}\,\mu\,\beta_2(x) + Z^2(1\leftrightarrow2,x\leftrightarrow x^{-1}) \,.
\end{align}
where $a\leftrightarrow b$ denotes replacement of $a$ with $b$.

The final low-energy contribution is given by
\begin{align}
\label{EL3intermediateAnt}
&\ E_{L3} = 
 -\frac{\alpha}{\pi}
\,\left(\frac{1}{3\,\varepsilon}+
\frac{5}{9}+ \frac23
\LZasquaredmmu\right)\,\nonumber \\
&\ \times\left(\frac{\mu}{m_1^4}  - \frac{(1-\varepsilon)}{m_1^2\,m_2}\right)
\left<\bigl(\vec{\nabla} V\bigr)^2\right> 
+ \frac{\alpha}{\pi}\,\frac{(Z\,\alpha)^6}{n^3}\,\mu\,\beta_3(x)\nonumber \\
&\  + Z^2(1\leftrightarrow2,x\leftrightarrow x^{-1})\,.
\end{align}

The middle-energy contribution for antiprotonic systems is obtained in a straightforward way as
\begin{align}
E_M = &\ \frac{\alpha}{2\,\pi\,m^2_1\,m_2}
\bigg[\frac{17}{9} - \frac{2}{3\,\varepsilon}-\frac43\big(\ln m_1\,r+\gamma\big)  \bigg](\vec\nabla V)^2 \nonumber \\ &\
+Z^2\,(1\leftrightarrow2)\,.
\end{align}

The only high-energy part that will contribute is given by $E_{H3}$, 
\begin{align}
E_{H3} = \frac{\alpha}{\pi}\bigg(\frac{1}{m_1^3} + \frac{Z^2}{m_2^3}\bigg)\bigg(\frac16 - \frac{1}{3\,\varepsilon}\bigg)(\vec\nabla V)^2\,.
\end{align}
The other terms go to  polarizability of both particles, and they are already included in the $\alpha^6\,\mu$ contribution in Ref.~\cite{zatorski:22}. 

After summing all the contributions, the singularities exactly cancel each other, which leads to
\begin{widetext}
\begin{align}
E^{(7)}_\mathrm{\bar p} = &\ \frac{\mu\,\alpha(Z\alpha)^2}{90\,\pi\,m_1^4\,m_2^2} \bigg\langle\frac{1}{r^4}
\bigg( 105\, m_1^2  
 + 170\, m_1\,m_2 + 68 \,m_2^2 - 60\,m_1 (m_1+m_2)\,\big(\ln(m_1\,r) + \gamma \big)  \nonumber \\
&\    + 60\,(m_1+m_2)^2\,\ln_1\bigg)\bigg\rangle  
+ \frac{\mu\,\alpha(Z\alpha)^6}{\pi\,n^3}\big\langle\beta^\mathrm{NS}(x) + \vec L\cdot\vec s_1\,\beta^\mathrm{S1}(x)
+\vec L\cdot\vec s_2\,\beta^\mathrm{S2}(x) \nonumber \\
&\ + \vec s_1\cdot\vec s_2\,\beta^\mathrm{SS}(x) + (L^i L^j)^{(2)}\,s_1^i \,s_2^j\,\beta^\mathrm{LL}(x) 
\big\rangle +Z^2\,(1\leftrightarrow2,x\leftrightarrow x^{-1})\,.
\end{align}
Evaluating the expectation values, we obtain
\begin{align}
E^{(7)}_\mathrm{\bar p} = &\  \frac{\mu\,\alpha(Z\alpha)^6}{\pi\,\Delta}\,\eta_1^2\,
\bigg[\frac83\,\eta_2\,\bigg(\frac{l(l+1)}{n^5} - \frac{3}{n^3}\bigg) 
\bigg(H_{2l-2} + H_{2l+3} - H_{n+l} + \ln\frac{n}{2\,Z\alpha\,\eta_1} \bigg) \nonumber \\
&\ + \frac83\, \bigg(\frac{3}{n^3} - \frac{l(l+1)}{n^5}\bigg) \,\ln_1 
+ \frac{1}{n^3}\bigg( \frac{70}{3} - \eta_1\,\frac{44}{3} + \eta_1^2\,\frac25\bigg)
-\frac{4\,\eta_2\,(2l+1)}{n^4} \nonumber \\
&\ + \frac{1}{n^5}\bigg(  l\,(l+1)\bigg( - 6 + \eta_1\,\frac{28}{9} - \eta_1^2\,\frac{2}{15}\bigg) + \eta_2\,\frac43 \bigg)  
 \bigg] 
 + \frac{\mu\,\alpha(Z\alpha)^6}{\pi\,n^3}\big\langle\beta^\mathrm{NS}(x) + \vec L\cdot\vec s_1\,\beta^\mathrm{S1}(x) \nonumber \\
&\ 
+\vec L\cdot\vec s_2\,\beta^\mathrm{S2}(x) 
+ \vec s_1\cdot\vec s_2\,\beta^\mathrm{SS}(x) + (L^i L^j)^{(2)}\,s_1^i \,s_2^j\,\beta^\mathrm{LL}(x) \big\rangle
  +Z^2\,(1\leftrightarrow2,x\leftrightarrow x^{-1})\,. \label{119}
\end{align}
\end{widetext}

 The final result for antiprotonic atoms is thus very simple and compact.

\section{Summary}

We have derived a complete $\alpha\,(Z\,\alpha)^6$ and  $Z^2\alpha\,(Z\,\alpha)^6$ one-loop self-energy correction 
to the energy levels of a two-body system with angular momentum $l>0$. The obtained results are valid for
constituent particles of arbitrary masses and spin 1/2, with the nucleus being either pointlike
or of extended-size. For $l=1$, the results are presented in Eqs. (\ref{81}-\ref{101}), while for $l>1$ they are in Eqs. (\ref{E2}-\ref{E21}), 
and these results are presented in Mathematica format in the Supplemental Material \cite{supp}.
For the case of positronium, the results for $l=1$ are presented in Eq.~(\ref{102}-\ref{108}),
and those for rotational states of antiprotonic atoms are in Eq.~(\ref{119}). For hydrogenlike atoms in the nonrecoil limit,
our results agree with the former calculation in the literature \cite{jentschura:05} in the case of a point nucleus.
We present also the first-order recoil correction in Eq.~(\ref{hydrrec}) for $l=1$ and in Eq.~(\ref{hydrrecl}) for $l>1$,
which to our knowledge have not yet been considered in the literature.

What  is yet unknown is the pure exchange contribution of order $(Z\,\alpha)^7$. Once it is completed,
we aim to perform numerical calculation of relativistic Bethe logarithms and 
the electron (muon) vacuum polarization contributions. This will eventually allow for
very accurate results for $l>0$ states of arbitrary two-body systems, including muonic and antiprotonic atoms.

Finally, we note that using the operator form of the $\alpha\,(Z\,\alpha)^6$ correction in Eq.~(\ref{E7rad ops}) we found a small mistake
in the previous calculation of a similar correction to He ionization energies, which we describe in detail in Appendix  B.


\appendix
\begin{widetext}
\section{Operators contributing to $E_{H2}$} \label{app:ops}

Individual first-order operators that come from the anomalous magnetic moment of the first particle are

\begin{align}
\delta H_{1} =&\ - \frac{Z\,\alpha}{4\, m_1^4}\,  \vec L\cdot\vec{s}_{1}\,\bigg( p^{2}\,\frac{1}{r^3}  + \frac{1}{r^3}\,p^{2}\bigg) \,, \label{34}\\
\delta H_{2} =& \ \frac{Z \alpha }{2\,m_{1}^{2} m_{2}^{2}}\,(g_2-1)\,
( \vec{s}_2 \times \vec{p})^i \bigg( \frac{\delta^{ij}}{r^3} - 3 \frac{r^{i} r^{j}}{r^5} 
+\frac{\delta^{ij}}{3}\,4\,\pi\,\delta^3(r) \bigg) ( \vec{s}_1 \times \vec{p})^j
+ \frac{Z\alpha}{24\,m_1^2}\,\Big( r_{E2}^2 + \frac{3}{4\,m_2^2}\Big) \,4\pi\,\nabla^2\,\delta^3(r)\nonumber \\
&\ 
+i\frac{Z\alpha}{6\,m_1^2}\biggl[ \,\frac{3\,(g_2-1)}{4\,m_2^2}\vec s_2 
+ \Big( r_{E2}^2 + \frac{3}{4\,m_2^2}\Big)\,\vec s_1\biggr]\cdot\vec p\times 4\pi\delta^3(r)\,\vec p \,, \label{36} \\
\delta H_3  = &\   \frac{Z\alpha}{4\,m_1\,m_2^{3}}\,\bigg(
	\bigg\{p^{2} ,\, \vec{s}_2 \times \vec{\nabla}_2 \cdot  \frac{ \vec{s}_{1} \times \vec{r}}{r^3}\bigg\}
	- \bigg\{ p^{2} ,\, \vec{p} \cdot \frac{ \vec{s}_{1} \times \vec{r}}{r^3}\bigg\}\bigg)
	+ \frac{1}{4\,m_1^3} \{\vec{p} \cdot \vec{\nabla}_{1}\times e_{1}\vec{\cal{A}}_{1}, \vec p\cdot \vec{s}_1 \}
\nonumber \\ & 
	+ \frac{Z\alpha\,(g_2-2)}{8\,m_1\,m_2^3} \bigg\{\vec{p} \times \vec{\nabla}_{2}\cdot\frac{ \vec{s}_{1} \times \vec{r}}{r^3}, \vec{p}\cdot \vec{s}_2 \bigg\}
-  \frac{Z\alpha}{8\, m_1^3\,m_2} 
\Big(i\,\vec s_1\cdot \vec p\times 4\pi\,\delta^3(r)\,\vec p - g_2\,\vec{s}_2 \times \vec{p}\, \,4\pi\,\delta^3(r)\,\vec{s}_1 \times \vec{p}\Big)
\nonumber \\
&\ + \frac{Z\alpha}{6\, m_1\,m_2} \bigg(g_2\,r_{M2}^{2} +\frac{3\,(g_2-2)}{4\,m_2^2}\bigg)
 \,\vec{s}_2 \times \vec{p} \,\,4\pi\,\delta^3(r)\,\vec{s}_1 \times \vec{p}\,, \label{39} \\
\delta H_{4} =& \ \frac{Z\alpha}{m_1\,m_{2}} \, e_{2}\vec{\cal{A}}_{2}\cdot \frac{ \vec{s}_{1} \times \vec{r}}{r^3} \, , \label{41}\\
\delta H_{5} = &\  \frac{1}{8\,m_1^3}\, \frac{(Z \alpha)^2}{r^4} \, , \\
\delta H_6 = &\ 
 \frac{(Z\alpha)^2\,(g_2-1)}{2\,m_1\, m_{2}^2} \, \frac{\vec{s}_{2} \times \vec r}{r^3} \cdot  \frac{ \vec{s}_{1} \times \vec{r}}{r^3} 
 -\frac{Z\alpha}{m_{1}^2} \, \vec{s}_{1} \cdot \frac{\vec r}{r^3} \times e_{1}\vec{\cal{A}}_1\, , \label{45} \\
\delta H_{7} =&\ \frac{Z \alpha}{4\, m_{1}\,m_{2}}\,\bigg( \,\bigg\{\bigg[ \bigg(\vec{s}_{1}\times \frac{\vec{r}}{r}\bigg)^{i},
	\, \frac{p^2}{2\,m_1} \bigg], \frac{i\,Z\alpha\, r^{i}}{r^3}\bigg\} 
	 +  \bigg\{\bigg[ \frac{p^2}{2\, m_2}, \bigg[ \bigg(\vec{s}_{1}\times \frac{\vec{r}}{r}\bigg)^{i}, \, \frac{p^2}{2\,m_1} \bigg] \bigg], p^i\bigg\}
	 \bigg)
	   \nonumber \\
	&\ - \frac{Z\, \alpha\,g_2}{8\, m_{1}^{2}\,m_{2}^{2}} \bigg[ p^2, \bigg[p^2, \, \vec{s}_{1}\cdot\vec{s}_{2}\,\frac{2}{3\,r} 
	+  s_{1}^{i}\,s_{2}^{j} 
	\frac{1}{2\,r}\bigg( \frac{r^i r^j}{r^2} - \frac{\delta^{ij}}{3} \bigg) \bigg] \bigg] \, . \label{49}\\
\delta H_8 =&\  -\frac{Z\alpha}{m_{1}^2} \; \vec{s}_1 \cdot \frac{\vec r}{r^3} \times e_{1}\vec{\cal{A}}_1
	+ \frac{i}{4\,m_{1}^3} \,
	\big[ \,\{e_{1}\vec{\mathcal{A}}_1, \,\vec{p} \times \vec{s}_{1}\}, p^2 \, \big]
+\frac{(Z\alpha)^2\,(g_2-1)}{2\,m_1\, m_{2}^2} \; \frac{\vec{s}_2 \times \vec r}{r^3} \cdot  \frac{ \vec{s}_{1} \times \vec{r}}{r^3} \nonumber \\ &\ 
	- \frac{i Z\alpha\,(g_2-1)}{8\,m_1\,m_{2}^3} \,
	\bigg[ \, \bigg\{ \frac{ \vec{s}_{1} \times \vec{r}}{r^3} ,\vec{p} \times \vec{s}_{2} \bigg\}, p^2 \, \bigg]	
	\, . \label{51}\\
\delta H_{9} = &\
\frac{Z\,\alpha}{6\,m_1\,m_2}\, \Bigl(r_{E2}^2-\frac{3\,(g_2-2)}{8\,m_2^2}\,\Bigr)\,
 i\,\vec s_1\cdot\vec p\times 4\,\pi\,\delta^{3}(r)\,\vec p 
 + \frac{Z\,\alpha}{16\,m_1^3\,m_2}\,\big(
 i\,g_2\,\vec s_2\cdot\vec p\times 4\,\pi\,\delta^{3}(r)\,\vec p 
 + 2\,\pi\, \vec\nabla^2\delta^{3}(r)\big) \,, \label{57}
\end{align}
\end{widetext}
where
\begin{align}
	e_1 {\cal{A}}^{i}_{1} =& - \frac{Z\alpha}{2\,r} \left( \delta^{ij} + \frac{r^{i} r^{j}}{r^2} \right) \frac{p_{2}^j}{m_2} -
	\frac{Z\,\alpha\,g_2}{2\,m_2} \frac{ \left(\vec{s}_{2} \times \vec{r}\right)^i}{r^3}\,, \label{29} 	\\
        e_2 {\cal{A}}^{i}_{2} =& - \frac{Z\alpha}{2\,r} \left( \delta^{ij} + \frac{r^{i} r^{j}}{r^2} \right) \frac{p_{1}^j}{m_1} 
       + \frac{Z\,\alpha}{m_1} \frac{ \left(\vec{s}_{1} \times \vec{r}\right)^i}{r^3}\, ,  \label{30}
\end{align}
are static vector potentials.

\section{Comparison with helium $\alpha^7$ radiative corrections}

We can compare our results with electron-electron operators derived for helium centroid triplet states in Ref.~\cite{patkos:21:rad}, given by the expression $E^{B}_\mathrm{SE}$ in Eq.~(156) of that work.
It can be transformed into the form
\begin{align}
&\ E_\mathrm{SE}^B =  \frac{\alpha^7}{\pi}\bigg\langle -\bigg(\frac{1039}{1350}+\frac{49}{45}\ln\left[\half\,\alpha^{-2}\right]\bigg)\,\vec p\,4\pi\,\delta^3(r)\,\vec p \nonumber \\
&\ + \bigg(\frac{403}{90} + 2\,\ln\left[\half\,\alpha^{-2}\right] - \frac43\,\ln r - \frac43\,\gamma - \frac23\,\ln 2\bigg)\frac{1}{r^4}\bigg\rangle\,.
\end{align}
This result can be checked against our two-body first-order operators derived here. We obtain it from the general result $E^{(7)}_\mathrm{rad}$ in Eq.~(\ref{E7rad ops})
by setting $g_2=2$, $r_{E2}^2 = r_{M2}^2= 0$, $m_1 = m_2 = 1$,
adding the corresponding result for the second particle where we make the exchange $(1\leftrightarrow2)$,
setting $\vec s_1\cdot\vec s_2 = 1/4$,
omitting fine structure and hyperfine structure tensor terms, and transforming into atomic units by $r\rightarrow r/\alpha$. We obtain
\begin{align}
&\ \tilde E_\mathrm{SE}^B =  \frac{\alpha^7}{\pi}\bigg\langle -\bigg(\frac{1039}{1350}+\frac{49}{45}\ln\left[\half\,\alpha^{-2}\right]\bigg)\,\vec p\,4\pi\,\delta^3(r)\,\vec p \nonumber \\
&\ + \bigg(\frac{851}{180} + 2\,\ln\left[\half\,\alpha^{-2}\right] - \frac43\,\ln r - \frac43\,\gamma - \frac23\,\ln 2\bigg)\frac{1}{r^4}\bigg\rangle\,.
\end{align}
We observe a discrepancy between these results. It can be traced to the contribution $E_2$ in Ref.~\cite{patkos:21:rad}, given by Eqs.~(102), (103), and (104). There is a missing
overall factor of 2 in this term, which would lead to an additional contribution in helium results equal to
\begin{equation}
\delta E = \frac{\alpha^7}{4\,\pi\,r^4}\,.
\end{equation}
Correcting for this mistake, we would get a perfect agreement between the two results. 
The numerical change from this correction amounts only to 2~kHz for the $2^3S$ state
and 3~kHz for the $2^3P$ state, and thus does not explain discrepancies for ionization energies \cite{He1, He2}.

\section{Derivation of identities} \label{app:identities}

To derive Eq.~(\ref{83}), we start with the identity
\begin{align}
&\ p^i\,[ p^i,[V,p^j]]\,p^j \nonumber \\
 = &\ p^2\,V\,p^2 + \frac12\,p^i\,[p^j,[V,p^j]]\,p^i  - \frac12\,\{p^2,\vec p\,\,V \vec p\,\}\nonumber \\
= &\ \frac14\,[p^2,[V,p^2]] + \frac12\,p^i\,[p^j,[V,p^j]]\,p^i  - \frac14\,\{p^2,[\vec p,[V, \vec p\,]]\}\,.
\end{align}
For states with $l>0$ the third term in the last equality vanishes. With the help of the expectation value identity
\begin{equation}
[p^2,[V,p^2]] = 4\,\mu(\vec \nabla V)^2\,,
\end{equation}
and relation
\begin{equation}
[ p^i,[V,p^j]] = \bigg[\frac{Z\alpha}{r^3}\bigg(\delta^{ij} - 3\,\frac{r^i\,r^j}{r^2}\bigg)\bigg]_\varepsilon + \frac{\delta^{ij}}{d}\,Z\alpha\,4\pi\,\delta^d(r)\,,
\end{equation}
with $d=3-2\,\varepsilon$, we arrive at Eq.~(\ref{83}).

We will also present the evaluation for the expectation value of the operator in the first term of Eq.~(\ref{36}),  
\begin{equation}
\bigg\langle( \vec{s}_2 \times \vec{p})^i \left( \frac{\delta^{ij}}{r^3} - 3 \frac{r^{i} r^{j}}{r^5} 
+\frac{\delta^{ij}}{3}\,4\,\pi\,\delta^3(r) \right) ( \vec{s}_1 \times \vec{p})^j\bigg\rangle\,.
\end{equation}
First, we need to isolate the traceless part of this operator, which is contracted with spin vectors.
The expectation value of the traceless part will be proportional to $\langle (L^i L^j)^{(2)}\,s_1^i s_2^j\rangle$,
while the trace part will result in terms involving $\langle \vec s_1\cdot\vec s_2\rangle$. For the non-local term we get
\begin{align}
 \bigg\langle( \vec{s}_2 \times \vec{p})^i  \bigg(  &\ \frac{\delta^{ij}}{r^3} - 3 \frac{r^{i} r^{j}}{r^5}  
\bigg)  ( \vec{s}_1 \times \vec{p})^j\bigg\rangle  \nonumber \\ 
 = &\  \big\langle\, \big( A\,(L^i L^j)^{(2)} + B\,\delta^{ij} \big)\,s_1^i s_2^j\big\rangle\,.
\end{align}
Coefficients $A$ and $B$ are obtained by projecting the expression on both sides of the equation, which is contracted with spin operators, either
to $(L^i L^j)^{(2)}$ or $\delta^{ij}$. After lengthy angular momentum algebra, this leads to
\begin{align}
A = &\ \frac{1}{(2l-1)(2l+3)}\,\bigg\langle\frac{10}{3}\,\vec p\,4\pi\,\delta^3(r)\,\vec p - \frac{12\,\mu\,E}{r^3} \nonumber \\ &\ - \frac{16\,\mu\,Z\alpha}{r^4}\bigg\rangle\,,\\
B = &\ -\frac13\,\bigg(\frac16\,\vec p\,4\pi\,\delta^3\,\vec p + \frac{\mu\,Z\alpha}{r^4}\bigg)\,.
\end{align}
For the local interaction part we would proceed in a similar way, leading to
\begin{align}
&\ \bigg\langle( \vec{s}_2 \times \vec{p})^i \,4\,\pi\,\delta^3(r) ( \vec{s}_1 \times \vec{p})^i\bigg\rangle
=  \langle \vec p\,4\pi\,\delta^3(r)\,\vec p\,\rangle\,\nonumber \\
&\ \times\bigg\langle (L^i L^j)^{(2)}\,s_1^i s_2^j+ \frac23\,\vec s_1\cdot\vec s_2\bigg\rangle\,.
\end{align}

\section{Expectation values of first-order operators} \label{app:formulas}

We employ the following identities to evaluate the expectation values with hydrogenic wave functions \cite{drakeLog,bethe}:
\begin{align}
\bigg\langle\frac{1}{r^3}\bigg\rangle = &\ \frac{2\,(\mu\,Z\alpha)^3}{l\,(l+1)\,(2l+1)\,n^3}\,\\
\bigg\langle\frac{1}{r^4}\bigg\rangle = &\ \frac{4\,(\mu\,Z\alpha)^4\,\big(3\,n^2 - l(l+1)\big)}{l\,(l+1)\,(2l-1)\,(2l+1)\,(2l+3)\,n^5}\,,\\
\bigg\langle\frac{\ln m_1\,r + \gamma}{r^4}\bigg\rangle = &\ \frac{(\mu\,Z\alpha)^4}{l\,(l+1)\,(2l-1)\,(2l+1)\,(2l+3)\,n^5} \nonumber \\
&\ \hspace{-5.0em} \times \bigg[
4\,\big(3\,n^2 - l(l+1)\big)\,\bigg( H_{2l-2} + H_{2l+3} - H_{n+l} \nonumber \\ 
&\ \hspace{-5.0em} + \ln\frac{n\,m_1}{2\,\mu\,Z\alpha} - \frac12\bigg) 
 - 2 \,\big( 1 - 3 \,(2l+1)\,n + 4\,n^2\big)\bigg]\,, \\
\langle\,\vec p\,4\pi\,\delta^3(r)\,\vec p \,\rangle = &\ \frac{4\,(\mu\,Z\alpha)^5}{3}\,\bigg(\frac{1}{n^3} - \frac{1}{n^5}\bigg)\,\delta_{l1}\,.
\end{align}

\section{General results for states with $l>1$} \label{app:results}

In this section, we will present the results for arbitrary angular momentum $l>1$. Defining
\begin{equation}\label{delta}
\Delta = l (l+1) (2l-1) (2l+1) (2l+3)\,,
\end{equation}
we obtain the following results for the coefficients in Eq.~(\ref{res}):

\begin{widetext}
\begin{align}
\mathcal{E}_\mathrm{NS} = &\
\frac{\mathcal{E}_\mathrm{NS}^{(3)}}{n^3} + \frac{\mathcal{E}_\mathrm{NS}^{(4)}}{n^4} + \frac{\mathcal{E}_\mathrm{NS}^{(5)}}{n^5}
+ \eta_1^2\,\eta_2\,\frac{8\,\big(l\,(l+1)-3\,n^2\big)}{3\,n^5}\big(H_{2l-2} + H_{2l+3} - H_{n+l} +\ln\frac{n}{2\,\eta_1\,Z\alpha}\big)
+\frac{\Delta}{n^3}\,\beta_\mathrm{NS}\,, \label{E2}
\\
\mathcal{E}_\mathrm{NS}^{(3)} = &\ \eta_1^2\bigg(8\,\ln_1 + \frac{113}{6} + \frac{3}{2l} - \frac{3}{2 (l+1)} + \frac{4}{(2l+1)^2} 
+ \eta_1\,\bigg( -\frac{79}{6} - \frac{3}{4l} + \frac{3}{4(l+1)} - \frac{2}{(2l+1)^2} \bigg)\nonumber \\
&\ + \eta_1^2\,\frac25\bigg) + \eta_1^2\,\eta_2^2\,\bigg( - \frac{3 }{16l} + \frac{3 }{16(l+1)} - \frac{3 }{16(2l-1)} - \frac{ 3 }{8(2l+1)^2}
+\frac{3 }{16(2l+3)}\bigg)\,g_2^2\,,\\
\mathcal{E}_\mathrm{NS}^{(4)} = &\ \eta_1^2\bigg[- \frac{3 (2l+1)}{4} + \frac{3}{(2l+1)} + \eta_2\bigg(-\frac{19(2l+1)}{4} + \frac{3}{(2l+1)}\bigg)\bigg]
- \eta_1^2\,\eta_2^2\,\frac{9}{16(2l+1)}\, g_2^2\,, \\
\mathcal{E}_\mathrm{NS}^{(5)} = &\ 
 \eta_1^2\, l(l+1)\,\bigg( -\frac83\,\ln_1
 - \frac{11}{2} + \frac{4}{3l} - \frac{4}{3(l+1)} + \eta_1\bigg(\frac{28}{9} - \frac{4}{3l} + \frac{4}{3(l+1)} \bigg)
 -\eta_1^2\,\frac{2}{15}\bigg)\,.
\end{align}
The following coefficient is
\begin{align}
\mathcal{E}_\mathrm{S1} = &\
\frac{\mathcal{E}_\mathrm{S1}^{(3)}}{n^3} + \frac{\mathcal{E}_\mathrm{S1}^{(4)}}{n^4} + \frac{\mathcal{E}_\mathrm{S1}^{(5)}}{n^5} + \frac{\Delta}{n^3}\,\beta_\mathrm{L1}\,, \\
\mathcal{E}_\mathrm{S1}^{(3)} = &\ \eta_1 \bigg[6 - \frac{3}{l} + \frac{3}{(l+1)} - \frac{8}{(2l+1)^2} - 6\,\eta_1\,\eta_2
+\bigg(\eta_1(\eta_1-2) + \eta_2^3\frac{g_2}{4}\bigg)\bigg(\frac{3}{2l^2} - \frac{13}{2l} + \frac{3}{2(l+1)^2} + \frac{13}{2(l+1)} \nonumber \\
&\ - \frac{16}{(2l+1)^2}\bigg)\bigg]  + \eta_1\,\eta_2^2\,g_2^2\bigg[\frac{3}{16l^2} - \frac{1}{16l} + \frac{3}{16(l+1)^2} + \frac{1}{16(l+1)}
+ \frac{3}{4(2l-1)} - \frac{1}{2(2l+1)^2} - \frac{3}{4(2l+3)} \nonumber \\
&\ + \eta_2\bigg(-\frac{9}{16l^2} + \frac{27}{16l} - \frac{9}{16(l+1)^2} - \frac{27}{16(l+1)} - \frac{3}{4(2l-1)} 
+\frac{9}{2(2l+1)^2} + \frac{3}{4(2l+3)}\bigg)\bigg]
\,,\\
\mathcal{E}_\mathrm{S1}^{(4)} = &\ \eta_1\bigg[3 - \frac{9}{2l} + 6l  - \frac{9}{2(l+1)} + \frac{12}{(2l+1)}
+\eta_2^2\bigg(1+\eta_2\frac{g_2}{4}\bigg)\bigg(\frac{9}{2l} + \frac{9}{2(l+1)} - \frac{24}{(2l+1)}\bigg)\bigg] \nonumber \\
&\ + \eta_1\,\eta_2^2\,g_2^2\bigg[\frac{9}{16l} + \frac{9}{16(l+1)} - \frac{3}{4(2l+1)} + \eta_2\bigg(
-\frac{27}{16l} - \frac{27}{16(l+1)} + \frac{27}{4(2l+1)}\bigg)\bigg]\,, \\
\mathcal{E}_\mathrm{S1}^{(5)} = &\ \eta_1\,l(l+1)\bigg[-8 + \frac{9}{2l} - \frac{9}{2(l+1)} +\eta_2\bigg(6 - \frac{3}{l} + \frac{3}{(l+1)}\bigg)
+\eta_2^2\bigg(-6 + \frac{9}{2l} - \frac{9}{2(l+1)}\bigg)\bigg] +\eta_1 \eta_2^3\frac{3}{8}\,(g_2-g_2^2)\,.
\end{align}
For coefficient $\mathcal{E}_\mathrm{S2}$ we obtain
\begin{align}
\mathcal{E}_\mathrm{S2} = &\
\frac{\mathcal{E}_\mathrm{S2}^{(3)}}{n^3} + \frac{\mathcal{E}_\mathrm{S2}^{(4)}}{n^4} + \frac{\mathcal{E}_\mathrm{S2}^{(5)}}{n^5} + \frac{\Delta}{n^3}\,\beta_\mathrm{S2}\,, \\
\mathcal{E}_\mathrm{S2}^{(3)} = &\ \eta_1^2\,\eta_2(3+\eta_2)\,g_2\bigg(-\frac{3}{8l^2} + \frac{13}{8l} - \frac{3}{8(l+1)^2}
-\frac{13}{8(l+1)} + \frac{4}{(2l+1)^2} \bigg) \nonumber \\
&\ + \eta_1^2\,\eta_2^2\,g_2^2\bigg(\frac{9}{16l^2} - \frac{27}{16l} + \frac{9}{16(l+1)^2} + \frac{27}{16(l+1)} + \frac{3}{4(2l-1)}
- \frac{9}{2(2l+1)^2} - \frac{3}{4(2l+3)}\bigg) \,, \\
\mathcal{E}_\mathrm{S2}^{(4)} = &\ \eta_1^2\,\eta_2\,g_2\bigg[-\frac{27}{8l} - \frac{27}{8(l+1)} + \frac{18}{(2l+1)}
+ \eta_2\bigg(-\frac{9}{8l} - \frac{9}{8(l+1)} + \frac{6}{(2l+1)}\bigg)\bigg]\nonumber \\
&\ + \eta_1^2\,\eta_2^2\,g_2^2\bigg(\frac{27}{16l} + \frac{27}{16(l+1)} - \frac{27}{4(2l+1)}\bigg)\,, \\
\mathcal{E}_\mathrm{S2}^{(5)} = &\ \eta_1^2\,\eta_2\,g_2\bigg(-\frac98 - \eta_2\,\frac38\bigg) + \eta_1^2\,\eta_2^2\,\frac{3}{8}\,g_2^2\,.
\end{align}
The scalar spin-spin coefficient is
\begin{align}
\mathcal{E}_\mathrm{SS} = &\
\frac{\mathcal{E}_\mathrm{SS}^{(3)}}{n^3} + \frac{\mathcal{E}_\mathrm{SS}^{(4)}}{n^4} + \frac{\mathcal{E}_\mathrm{SS}^{(5)}}{n^5} + \frac{\Delta}{n^3}\,\beta_\mathrm{SS}\,, \\
\mathcal{E}_\mathrm{SS}^{(3)} = & \eta_1\,\eta_2^2\,\bigg(2-\frac{1}{l} + \frac{1}{(l+1)} - \frac{8}{3(2l+1)^2}\bigg)
+\eta_1\,\eta_2\,g_2\bigg(-\frac52 + \frac{1}{l} - \frac{1}{(l+1)} + \frac{8}{3(2l+1)^2}\bigg) \nonumber \\ &\
 + \eta_1^2\,\eta_2^2\,g_2^2\bigg(-\frac{1}{4l} + \frac{1}{4(l+1)} - \frac{1}{4(2l-1)} - \frac{1}{2(2l+1)^2} + \frac{1}{4(2l+3)}\bigg)\,, \\
\mathcal{E}_\mathrm{SS}^{(4)} = &\ \eta_1\,\eta_2\,(\eta_2 - g_2)\bigg(2l+1 - \frac{4}{(2l+1)}\bigg)
 -\eta_1^2\,\eta_2^2\frac{3 }{4(2l+1)}\,g_2^2\,, \\
\mathcal{E}_\mathrm{SS}^{(5)} = &\ \eta_1\,\eta_2\,\frac{g_2 l(l+1)}{6} \,.
\end{align}
Finally, for the tensor spin-spin coefficient we obtain
\begin{align}
\mathcal{E}_\mathrm{LL} = &\
\frac{\mathcal{E}_\mathrm{LL}^{(3)}}{n^3} + \frac{\mathcal{E}_\mathrm{LL}^{(4)}}{n^4} + \frac{\mathcal{E}_\mathrm{LL}^{(5)}}{n^5} + \frac{\Delta}{n^3}\,\beta_\mathrm{LL}\,, \\
\mathcal{E}_\mathrm{LL}^{(3)} = & \eta_1\,\eta_2^2\bigg(-\frac{3}{l^2} + \frac{13}{l} - \frac{3}{(l+1)^2} - \frac{13}{(l+1)}
+ \frac{9}{(2l-1)} + \frac{32}{(2l+1)^2} - \frac{9}{(2l+3)}\bigg)
+ \eta_1\,\eta_2\,g_2\bigg[\eta_2^2 \left(\frac{9}{2 l^2}+\frac{27}{2 (l+1)} \right. \nonumber \\
&\ \left. +\frac{9}{2 l-1}-\frac{9}{2 l+3}+\frac{9}{2 (l+1)^2}-\frac{36}{(2 l+1)^2}-\frac{27}{2 l}\right) 
+\eta_2 \left(\frac{9}{2
    l^2}+\frac{27}{2 (l+1)}+\frac{18}{2 l-1}-\frac{18}{2 l+3} \right.\nonumber \\
&\ \left. +\frac{9}{2 (l+1)^2}-\frac{36}{(2 l+1)^2}-\frac{27}{2 l}\right) 
    -\frac{15}{4 l^2}+\frac{17}{4 l}-\frac{17}{4 (l+1)}-\frac{24}{2
    l-1}+\frac{24}{2 l+3}-\frac{15}{4 (l+1)^2}+\frac{16}{(2 l+1)^2}
    \bigg] \nonumber \\
&\ + \eta_1\,\eta_2^2\,g_2^2\bigg[
\frac{3}{2 l^2}-\frac{1}{2 l}+\frac{1}{2 (l+1)}+\frac{1}{2 (2 l-1)}-\frac{1}{2 (2 l+3)}+\frac{3}{2 (l+1)^2}-\frac{3}{(2 l-1)^2}-\frac{6}{(2 l+1)^2}-\frac{3}{(2 l+3)^2} \nonumber \\
&\ + \eta_2 \left(-\frac{15}{4 l^2}-\frac{29}{4 (l+1)}-\frac{29}{4 (2 l-1)}+\frac{29}{4 (2 l+3)}-\frac{15}{4 (l+1)^2}+\frac{3}{(2 l-1)^2}+\frac{24}{(2 l+1)^2}+\frac{3}{(2
    l+3)^2}+\frac{29}{4 l}\right)\bigg]\,, \\
\mathcal{E}_\mathrm{LL}^{(4)} = &\
\eta_1\,\eta_2\,g_2\bigg[-\frac{45}{4l} - \frac{45}{4(l+1)} + \frac{24}{(2l+1)} + \eta_2(1+\eta_2)\bigg(\frac{27}{2l}
+ \frac{27}{2(l+1)} - \frac{54}{(2l+1)}\bigg)\bigg]\nonumber \\
&\ 
+\eta_1\,\eta_2^2\,g_2^2 \bigg[-\frac{27}{4l} - \frac{27}{4(l+1)}
+ \frac{27}{(2l+1)} + \eta_1\bigg(\frac{45}{4l} + \frac{45}{4(l+1)} - \frac{9}{2(2l-1)} - \frac{36}{(2l+1)}
- \frac{9}{2(2l+3)}\bigg)\bigg] \nonumber \\
&\ +  \eta_1\,\eta_2^2\bigg(-\frac{9}{l} -\frac{9}{(l+1)} + \frac{48}{(2l+1)}\bigg)\,, \\
\mathcal{E}_\mathrm{LL}^{(5)} = &\ \eta_1\,\eta_2\,g_2\bigg[\eta_2^2 \left(\frac{9}{4 (2 l+3)}-\frac{9}{4 (2 l-1)}+6\right)+\eta_2 \left(\frac{9}{2 (2 l+3)}-\frac{9}{2 (2 l-1)}-6\right)+\frac{6}{2 l-1}-\frac{6}{2 l+3}+\frac{47}{4}\bigg] \nonumber \\
&\ + \eta_1\,\eta_2^2\,(4\,\eta_2 -1)\,g_2^2\bigg(\frac{9}{16(2l-1)} - \frac{9}{16(2l+3)}\bigg)+ \eta_1\,\eta_2^2\bigg(-9 - \frac{9}{4(2l-1)} + \frac{9}{4(2l+3)}\bigg)\,. \label{E21}
\end{align}

Further, for the positronium atom
$l>1$ states we obtain 
\begin{align}\label{respos}
E^{(7)}_\mathrm{pos} = &\ \frac{m\,\alpha(Z\alpha)^6}{\pi\,\Delta}\big\langle \mathcal{E}^\mathrm{pos}_\mathrm{NS} + \vec L\cdot(\vec s_1 + \vec s_2)\,\mathcal{E}^\mathrm{pos}_\mathrm{LS}
+ \vec s_1\cdot\vec s_2\,\mathcal{E}^\mathrm{pos}_\mathrm{SS}
+ (L^i L^j)^{(2)}\,s_1^i s_2^j\,\mathcal{E}^\mathrm{pos}_\mathrm{LL}
\big\rangle \,,
\end{align}
where
\begin{align}
\mathcal{E}^\mathrm{pos}_\mathrm{NS} = &\ \frac{1}{n^3}\bigg(\frac{247 }{80} + \frac{15}{64l} - \frac{15}{64(l+1)} -\frac{3}{64(2l-1)}
+\frac{21}{32(2l+1)^2}+\frac{3}{64(2l+3)}\bigg)
+ \frac{1}{n^4}\bigg(-\frac{25}{32} - \frac{25l}{16} + \frac{63}{64(2l+1)}\bigg) \nonumber \\
&\ + \frac{1}{n^5}\bigg(\frac16 - \frac{179 l (l+1)  }{180}\bigg) 
+ \frac{\big( 3n^2 - l(l+1)\big)}{3n^5}\bigg(2\LZasquaredpos - H_{2l-2} - H_{2l+3} + H_{n+l} - \ln\frac{n}{ Z\alpha}\bigg)
  + \frac{\Delta}{n^3}\,\beta_\mathrm{NS}(1)
\,, \\
\mathcal{E}^\mathrm{pos}_\mathrm{LS} = &\ \frac{1}{n^3}\bigg(\frac98 - \frac{3}{8l^2} + \frac{17}{16l} - \frac{3}{8(l+1)^2} - \frac{17}{16(l+1)}
+\frac{3}{16(2l-1)} + \frac{19}{8(2l+1)^2} - \frac{3}{16(2l+3)}\bigg)\nonumber \\
&\ + \frac{1}{n^4}\bigg(\frac34 - \frac{9}{8l} + \frac{3l}{2} - \frac{9}{8(l+1)} + \frac{57}{16(2l+1)}\bigg)
+\frac{1}{n^5}\bigg(\frac{57}{64} - \frac{13l(l+1)}{8}\bigg) + \frac{\Delta}{n^3}\,\beta_\mathrm{LS}(1)\,,\nonumber \\
\mathcal{E}^\mathrm{pos}_\mathrm{SS} = &\ \frac{1}{n^3}\bigg(-1 + \frac{5}{16l} - \frac{5}{16(l+1)} - \frac{1}{16(2l-1)}
+\frac{7}{8(2l+1)^2} + \frac{1}{16(2l+3)}\bigg)
+\frac{1}{n^4}\bigg(-\frac38 - \frac{3l}{4}+ \frac{21}{16(2l+1)}\bigg)\nonumber \\
&\ + \frac{l(l+1)}{12\,n^5} + \frac{\Delta}{n^3}\,\beta_\mathrm{SS}(1)\,,\\
\mathcal{E}^\mathrm{pos}_\mathrm{LL} = &\ \frac{1}{n^3}\bigg(-\frac{3}{4l^2} + \frac{1}{4l} - \frac{3}{4(l+1)^2} - \frac{1}{4(l+1)} - \frac{3}{4(2l-1)^2}
- \frac{109}{16(2l-1)} + \frac{3}{2(2l+1)^2} - \frac{3}{4(2l+3)^2} + \frac{109}{16(2l+3)}\bigg)\nonumber \\
&\ + \frac{1}{n^4}\bigg(-\frac{9}{4l} - \frac{9}{4(l+1)} - \frac{9}{8(2l-1)} + \frac{9}{4(2l+1)} - \frac{9}{8(2l+3)}\bigg)
+\frac{1}{n^5}\bigg(4 + \frac{51}{32(2l-1)} - \frac{51}{32(2l+3)}\bigg) \nonumber \\ &\ + \frac{\Delta}{n^3}\,\beta_\mathrm{LL}(1)\,.
\end{align}

For hydrogenlike atoms with $l>1$, in the limit of an infinitely heavy nucleus, we get the result
\begin{align}
E^{(7,0)}_{\rm hydr} = &\ \frac{m_1\,\alpha(Z\alpha)^6}{\pi\,\Delta}
\bigg[\frac{1}{n^3}\bigg(\frac{91}{15} + \frac{3}{4l} - \frac{3}{4(l+1)} + \frac{2}{(2l+1)^2}\bigg)
+\frac{1}{n^4}\bigg(-\frac34  - \frac{3l}{2} + \frac{3}{(2l+1)} \bigg) - \frac{227\,l(l+1)}{90\,n^5} \nonumber \\
&\ +  \frac{8\,\big(3n^2 - l(l+1)\big)}{3\,n^5}\LZasquared
+\langle\vec L\cdot\vec s_1\rangle\,\bigg(\frac{1}{n^3}\bigg(6 - \frac{3}{2l^2} + \frac{7}{2l} - \frac{3}{2(l+1)^2} - \frac{7}{2(l+1)} + \frac{8}{(2l+1)^2}\bigg) \nonumber \\
&\ + \frac{1}{n^4}\bigg(3 - \frac{9}{2l} + 6l - \frac{9}{2(l+1)} + \frac{12}{(2l+1)}\bigg)
+\frac{1}{n^5}\bigg(\frac92 - 8 l(l+1)\bigg)\bigg)
+  \frac{1}{n^3}\,\big(\beta_\mathrm{NS}(0) + \langle\vec L\cdot\vec s_1\rangle \,\beta_\mathrm{S1}(0)\big)
\bigg]  \,,
\end{align}
and for the leading recoil correction we get
\begin{align}\label{hydrrecl}
E^{(7,1)}_{\rm hydr} = & \frac{m_1^2\,\alpha(Z\alpha)^6}{\pi\,m_2\,\Delta} \big\langle \mathcal{E}^{(7,1)}_\mathrm{NS}
+ \vec L\cdot\vec s_1\,\mathcal{E}^{(7,1)}_\mathrm{S1}
+\vec L\cdot\vec s_2\,\mathcal{E}^{(7,1)}_\mathrm{S2}
+\vec s_1\cdot\vec s_2\,\mathcal{E}^{(7,1)}_\mathrm{SS}
+(L^i L^j)^{(2)} s_1^i s_2^j\,\mathcal{E}^{(7,1)}_\mathrm{LL}\big\rangle   \,, \\
\mathcal{E}^{(7,1)}_\mathrm{NS} = &\ \frac{1}{n^3}\bigg(\frac{13}{6}-\frac{3}{2l} + \frac{3}{(2l+1)} - \frac{4}{(2l+1)^2}\bigg)
-\frac{1}{n^4}\bigg(\frac52  + 5l + \frac{6}{(2l+1)}\bigg)+ \frac{1}{n^5}\bigg(\frac43 + \frac{37l(l+1)}{18}\bigg) \nonumber \\
&\ + \frac{8\big(l(l+1) - 3 n^2\big)}{3\,n^5}\,\bigg(3\,\LZasquared + H_{2l-2} + H_{2l+3} - H_{n+l} + \ln\frac{n}{2 Z\alpha}\bigg) 
+ \frac{\Delta}{n^3}\,(\beta_\mathrm{NS}^{(1)}-\beta_\mathrm{NS}^{(0)})\,\\
\mathcal{E}^{(7,1)}_\mathrm{S1} = &\ \frac{1}{n^3}\bigg(-18 + \frac{3}{l^2}
 - \frac{7}{l} + \frac{3}{(l+1)^2} + \frac{7}{(l+1)} - \frac{16}{(2l+1)^2}\bigg)
+\frac{1}{n^4}\bigg(-6 + \frac{9}{l} - 12l +\frac{9}{(l+1)} - \frac{24}{(2l+1)}\bigg) \nonumber \\
&\ + \frac{1}{n^5}\bigg(-12 + 22\, l(l+1)\bigg) + \frac{\Delta}{n^3}\,(\beta_\mathrm{S1}^{(1)} - \beta_\mathrm{S1}^{(0)})\nonumber \\
\mathcal{E}^{(7,1)}_\mathrm{S2} = &\ g_2\bigg[\frac{1}{n^3}\bigg(-\frac{9}{8l^2} + \frac{39}{8l} - \frac{9}{8(l+1)^2} - \frac{39}{8(l+1)}
+ \frac{12}{(2l+1)^2}\bigg) + \frac{1}{n^4}\bigg(-\frac{27}{8l} -\frac{27}{8(l+1)} + \frac{18}{(2l+1)}\bigg) - \frac{9}{8\,n^5}\bigg] \nonumber \\
&\ + \frac{\Delta}{n^3}\,\beta_\mathrm{S2}^{(1)}\,, \\
\mathcal{E}^{(7,1)}_\mathrm{SS} = &\ g_2\bigg[\frac{1}{n^3}\bigg(-\frac52 + \frac{1}{l} - \frac{1}{(l+1)} + \frac{8}{3(2l+1)^2}\bigg)
+\frac{1}{n^4}\bigg(-1 - 2 l + \frac{4}{(2l+1)}\bigg) + \frac{l(l+1)}{6\,n^5}\bigg] + \frac{\Delta}{n^3}\,\beta_\mathrm{SS}^{(1)}\,,\\
\mathcal{E}^{(7,1)}_\mathrm{LL} = &\ g_2\bigg[\frac{1}{n^3}\bigg(-\frac{15}{4l^2} + \frac{17}{4l} - \frac{15}{4(l+1)^2} - \frac{17}{4(l+1)} - \frac{24}{(2l-1)}
+ \frac{16}{(2l+1)^2} + \frac{24}{(2l+3)}\bigg)
+\frac{1}{n^4}\bigg(-\frac{45}{4l}-\frac{45}{4(l+1)} \nonumber \\
&\ + \frac{24}{(2l+1)}\bigg) + \frac{1}{n^5}\bigg(\frac{47}{4} + \frac{6}{(2l-1)} - \frac{6}{(2l+3)}\bigg)\bigg] + \frac{\Delta}{n^3}\,\beta_\mathrm{LL}^{(1)}\,.
\end{align}
\end{widetext}

\end{document}